\documentclass[showpacs,aps,superscriptaddress,pra,twocolumn,floatfix,nofootinbib]{revtex4-2}

\usepackage{dcolumn}% Align table columns on decimal point
\usepackage{bm}% bold math
\usepackage[utf8]{inputenc}
\usepackage[T1]{fontenc}
\usepackage{etoolbox}
\usepackage{graphicx}
\usepackage{amsmath}
\usepackage{amssymb}
\usepackage{tikz}
\usepackage{xcolor}

\definecolor{darkgreen}{RGB}{0,100,0}
\definecolor{darkmagenta}{RGB}{128,0,128}
\definecolor{darksea}{RGB}{0,105,148}

\usepackage{stmaryrd}
\usepackage{xspace}
\usepackage{booktabs}
\usepackage{float}
\usepackage{url}
\usepackage{nicefrac}
\usepackage{soul}
\usepackage{wasysym}
\usepackage{multirow}

\usepackage[colorlinks=true,breaklinks=true]{hyperref}

\hypersetup{
    linkcolor=red,
    citecolor=blue,
    urlcolor=blue
}

\begin{document}

\title{Structure and information measures of few-electron systems under a spherically symmetric Gaussian potential within a density functional approach}

\author{Raveena Arya}
\affiliation{Department of Chemical Sciences, Indian Institute of Science Education and Research (IISER) Kolkata, Mohanpur 741246, Nadia, India.
}

\author{Santanu Mondal}
\email[]{sansantanu1991@gmail.com}
\affiliation{Instituto de Ciencias Fisicas, Universidad Nacional Autonoma de Mexico, Av. Universidad S/N, Col. Chamilpa, Cuernavaca, Morelos, 62210, México
}

\author{Amlan K. Roy}
\email[]{akroy@iiserkol.ac.in, akroy6k@gmail.com}
\affiliation{Department of Chemical Sciences, Indian Institute of Science Education and Research (IISER) Kolkata, Mohanpur 741246, Nadia, India.
}

\begin{abstract} 
Energies of H, He-like ($Z=2-18$) ions, Li, and Be are investigated under a spherically symmetric Gaussian potential through a density functional formalism. The radial Kohn-Sham equation has been solved by invoking a work function-based exchange potential. The effect of electron correlation is analyzed by incorporating two functionals: a local parameterized Wigner functional and a non-linear gradient- and Laplacian-dependent Lee-Yang-Parr (LYP) functional. The generalized pseudospectral method is employed to provide accurate numerical eigenfunctions and eigenvalues. This allows \textit{nonuniform}, \textit{optimal} spatial discretization fulfilling the Dirichlet boundary conditions. This work demonstrates a possible manipulation of energy by controlling dot parameters. Apart from ground states, exploratory results are also reported for low-lying excited state $1s2s$ ($^{1,3}S$) of He atom. Companion calculations are also performed for various information-theoretic measures, such as Shannon entropy in position ($S_{r}$), momentum ($S_{p}$) spaces, and Fisher information in position space ($I_{r}$). The behavior of correlation functionals in presence of Gaussian potential is examined critically. We find that energy increases, $S_{r}$ exhibits minima, while $S_{p}$, $I_{r}$ attain maxima for a decrease in the width of potential, whereas an increase in potential depth further amplifies these effects across all properties. The Fisher-Shannon plane reveals a progressive localization as well as the compression of electronic density, and thereby indicates a weakening of relative electron-correlation effects. In the Collin's conjecture, it gives rise to a non-linear loop-like feature. Much of the results are presented here for the first time. 
 
\end{abstract}
\maketitle

\section{Introduction}\label{sec:intro}
Quantum confinement has long been a subject of significant interest in various branches of physics and chemistry, as it enables additional manipulation and control of atomic properties through confinement parameters. 
As a consequence, the system begins to exhibit modifications in its electronic and optical properties as compared to its free counterpart. 
Moreover, the evolution of modern-day technology has also made it feasible to design and produce such confinement through semiconductor nano-structures with distinct reduced dimensions, such as quantum dots (QDs), quantum wells, quantum well wires, and super-lattices \cite{jiban,alivisatos,ricky,kvon,diroll}. 

The QD \cite{Johnson,tarucha}, a zero-dimensional quantum structure, represents a nano-scale conducting system with wide-spread applications in various fields, including electronics, energy technology, quantum information science, biotechnology, modeling molecules with tunable bonds, LEDs, semiconductors, transistors, diode lasers, solar cells, medical imaging, and opto-electronics \cite{singhal, ding,sun25,moungi,kirmani}. 
The progress of micro-fabrication technologies has also made the realization of QDs feasible. 
Techniques such as molecular beam epitaxy and metal-organic chemical vapor 
deposition are widely employed to grow QDs with precisely controlled nano-scale dimensions \cite{Hao2019,nano12}. 
In principle, QDs can confine any number of electrons in all three spatial dimensions. 
The confinement of such electrons in QDs results in quantized bound states analogous to atomic orbitals. 
This is why often QDs are termed as \textit{artificial} atoms \cite{Ashoori1996}. 
Nevertheless, the resulting energy spectra, electron densities, and excitation dynamics are profoundly influenced by the confinement potential, making them distinct from those in real atoms.
From a theoretical perspective, various types of model potentials such as harmonic, Gaussian, finite oscillator, Woods-Saxon, P\"oschl-Teller, and Tietz type potential \cite{mondal,arya2025,Pandey2004,boya,gharaati09,winkler2004,Kimani2008,saha16,cari,khor14} have been proposed to describe such systems.

%%%%%%%%%%%%%%%%%%%%%%%%%%%%%%%%%%%%%%%%%%%%%%%%%
In the present work, confinement is represented by a \textit{Gaussian potential} (GP) to model atoms/ions within a spherical QD. 
It provides a soft, continuous boundary and has been widely used to describe effective interactions in nuclear reactions \cite{buck1977}, molecular vibrational and chemical transformation processes \cite{gribov2018,gribov2018a}, and atoms confined within fullerene cages \cite{nasc2011,lin_2012,lin2013}. 
For instance, the autoionizing resonances in a spherical Gaussian-shaped two-electron QD was probed using complex scaled direct diagonalization method \cite{genkin10}. 
Multipole transition oscillator strengths and static polarizabilities of hydrogenic impurities were calculated using a Bernstein polynomial-based basis-expansion technique \cite{lumb15}. 
Further, the ground-state energy of Be atom using a numerical, parameterized optimized effective potential method was analyzed in the GP confinement \cite{arias21}. 
The H-like donor impurity was examined using a variational method \cite{kachu24,semina16}, to evaluate transition energy, magnetic moment, and susceptibilities. 
Also, GP is used to simulate a fullerene cage environment in which structural properties and ionization dynamics of He atom confined within C$_{60}$ cages \cite{chen26} were investigated. 
Energetics of He atom in a GP has been reported employing an exact diagonalization method \cite{xie08}, using variational method in terms of Hylleraas-type basis functions \cite{laughlin09}. Two-electron Gaussian QD in excited state 
has also been studied in the absence of attraction potential \cite{sen21}. 

The characteristics of atoms/molecules are often examined through information measures like Shannon entropy ($S$) \cite{shannon48} and Fisher information ($I$) \cite{Fisher25}. They serve as valuable tools for quantifying uncertainty, randomness, disorder, localization, and electron correlation effects in such systems \cite{majumdar20,arya2025}. 
The present work is devoted to a detailed analysis of both structural properties and information measures of by tuning the parameters (depth $V_{0}$ and width $R$) of confining GP simulating a QD environment, having a potential of the form:
     \begin{equation}
\label{eq:1}
    V_{\texttt{conf}}(r) = V_{0}\left[1-\exp\left( \frac{-r^{2}}{R^{2}} \right)\right], 
\end{equation}
where $V_{0}$ is height of potential well and $R$ signifies confining radius.
Evidently, at the center of a QD, the potential behaves almost parabolically, while at the boundaries it changes smoothly, yielding a confinement of finite range and depth. As already mentioned, an atom within GP has been published in the literature in limited manner, mostly in terms of energetics in ground states of two electrons. But there is a genuine lack of works for more than 
two electrons, excited states and information theoretical studies. Thus specifically, we evaluate ground- and excited-state energies, and information entropies such as $S_{r}$, $S_{p}$, and $I_{r}$ 
for atoms/ions confined within a GP. 

This is achieved by making use of a general time-independent density functional theory (DFT)-based scheme, that is applicable for an arbitrary state of an atom/ion. This incorporates a physically motivated work-function-based exchange potential \cite{sahini_1990,sahni1992}. 
Correlation effects are treated using two functionals, $viz.,$ a simple, local, parametrized Wigner-type and a nonlinear Lee-Yang-Parr (LYP) functional. 
This is a basis-set free approach, which has demonstrated considerable success in describing static and dynamic properties in atomic and molecular problems \cite{roy97,roy2002,roy04jpbhollow,roy05jpbhollow,roy2007,majumdar20,majumdar2021,majumdar2021a,arya2025}. 
The resulting Kohn-Sham (KS) equation is constructed with the chosen XC functional and solved using the accurate and efficient generalized pseudo-spectral (GPS) method with Dirichlet boundary condition. 
The self-consistent field eigen functions thus obtained from the solution, results in the $r$-space density; the latter, in turn, provides the $r$-space $S_r$ and $I_r$. The Fourier transform of $r$-space orbitals gives the corresponding $p$-space orbitals, from which $p$-space densities are obtained. 
The latter then offers $p$-space entropies. 

The manuscript is organized as follows: A brief introduction of the methodology is discussed in next section, Sec.~\ref{sec:method}. 
The details of results are outlined in Sec.~\ref{sec:results} in which Sec.~\ref{sec:energy} describes the ground states of one-, two-, three- and four-electron systems in a GP. 
It is pertinent to mention that the confined H atom in a GP has already been investigated in earlier studies \cite{lumb15,liu24,zhou25,ma25}. 
Thus we do not delve in to the details of H atom. 
As He is the simplest example to understand the correlation effect in many-electron system, our main goal is to pursue He-like ($Z=2-18$) systems in presence of a GP. 
To investigate the correlation effect, correlation energy w.r.t. tuning parameters ($R, V_{0}$) and $Z$ are reported. 
Next, to extend it beyond two electrons, the case of Li and Be in this potential have been studied. 
Additionally, the low-lying excited states, 1s2s ($^3$S and $^1$S) for He atom 
are also considered. 
In the remaining sections, our focus remains centered mostly on He-like ions ($Z=2-18$). 
Next, in Sec.~\ref{sec:Shannon}, the calculated $S_r$ and $S_p$ in terms of $V_{0}$ and $R$ for the He-isoelectronic series, to gain insight in to the delocalization or uncertainty. 
Then Sec.~\ref{sec:fisher} discusses $I_r$, similar to that done for $S_r$ and $S_p$. 
Thereafter, Fisher-Shannon information plane has been investigated in Sec.~\ref {sec:fisher-shannon}, where $S_{r}$ (global) and $I_{r}$ (local) measures can be studied simultaneously in the GP. 
Lastly, we have revisited the Collin's conjecture in Sec.~\ref{sec:collins}, where the trend between correlated Shannon entropies in $r, p$ and combined spaces ($S_{\texttt{r}_\texttt{corr}}$, $S_{\texttt{p}_\texttt{corr}}$, $S_{\texttt{t}_\texttt{corr}}$) w.r.t. to correlated energy, $E_{\texttt{corr}}$, have been demonstrated. Finally, a few conclusions are drawn in Sec.~\ref{sec:conc}. 

\section{Methodology}\label{sec:method}
Here, we provide a brief overview of the DFT method for an arbitrary atom/ion confined by a model GP to describe the QD. 
The method is well established \cite{roy97,roy2002,roy04jpbhollow,roy05jpbhollow,roy2007,majumdar20,majumdar2021,majumdar2021a,arya2025}, as has been demonstrated for a variety of calculations in both ground and excited states in atoms--including singly, doubly, and triply excited states; valence and core excitations; low- and high-lying states; auto-ionizing, hollow, and doubly hollow states; high-lying Rydberg states; and satellite states, as well as some recent studies in quantum confinement of atoms/ions in different physical situations. 
The starting point is the non-relativistic, single-particle, time-independent KS equation under the desired GP confinement as,
\begin{equation}
\label{eq:2}
    \hat{H}(\vec{r})\psi_i(\vec{r})=\epsilon_{i}(\vec{r})\psi_{i}(\vec{r})\,,
\end{equation}
where $\hat{H}$ refers to the effective KS Hamiltonian, given by,
\begin{eqnarray}
\label{eq:3}
    \hat{H}(\vec{r}) =  -\frac{1}{2}\nabla^{2}+v_{eff}(\vec{r})\,,
\end{eqnarray}
where, the effective potential
\begin{eqnarray}
    v_{eff}(\vec{r}) =  v_{ne}(\vec{r}) +\int \frac{\rho(\vec{r}^{\prime})}{|\vec{r} - \vec{r}^{\prime}|}
 \mathrm{d}\vec{r}^{\prime}+\frac{\delta E_{xc}[\rho(\vec{r})]}{\delta \rho(\vec{r})} + v_{conf}(\vec{r}) \nonumber\,.
\end{eqnarray}
Here $v_{ne}(\vec{r})$ denotes the external electron-nuclear attraction potential, whereas second and third terms in right-hand side represent classical Coulomb (Hartree) repulsion and XC potentials, respectively. 
The QD in question is represented by a model potential, given in Eq.~\eqref{eq:1}.
An accurate work-function-based exchange potential, $v_x(\vec{r})$, is employed \cite{sahini_1990,sahni1992}, where the exchange energy is obtained by treating it as an interaction between an electron at $\vec{r}$ and its associated Fermi-Coulomb hole charge density, $\rho_{x}(\vec{r},\vec{r}^{\prime})$, at $\vec{r}^{\prime}$. 
The explicit expression of exchange energy is given by \cite{sahini_1990,sahni1992}, 
\begin{equation}
\label{eq:4}
 E_{x}[\rho(\vec{r})] = \frac{1}{2}\int\int\frac{\rho(\vec{r})\rho_{x}(\vec{r},\vec{r}^{\prime})}{|\vec{r} - \vec{r}^{\prime}|}
\mathrm{d}\vec{r}\mathrm{d}\vec{r}^{\prime}\,.
\end{equation}
The corresponding exchange potential $v_{x}(\vec{r})$, 
\begin{equation}
\label{eq:5}
v_{x} (\vec{r})  = -\int_{\infty}^{r} \mathcal{E}_{x}(\vec{r}) \cdot \mathrm{d} \vec{l} \,, 
\end{equation}
then can be interpreted as work required to bring an electron from infinity to a point $\vec{r}$ in the electric field generated by its Fermi-Coulomb hole charge density; the latter can be expressed as, 
\begin{equation}
\label{eq:6}
\mathcal{E}_{x}(\vec{r}) = \int\frac{\rho_{x}(\vec{r},\vec{r}^{\prime})(\vec{r} - \vec{r}^{\prime})} {|\vec{r} - \vec{r}^{\prime}|^{3}} \mathrm{d}{\vec{r}^{'}}\,.
\end{equation}
The correlation effect is incorporated through (i) a simple Wigner \cite{brual1978} and (ii) a slighly more involved LYP \cite{Lee_Parr_1988} functional.

\setlength{\tabcolsep}{6pt}
\begin{table}[!th]
\centering
\caption{Ground-state energy of He in GP, for different $V_{0}$ and $R$. All quantities are in a.u.}
\begin{tabular}{ccrrr}
\hline\hline
$V_{0}$ &$R$ & \multicolumn{1}{c}{$E_{\texttt{XX}}$} & \multicolumn{1}{c}{$E_{\texttt{XC-WIG}}$} & \multicolumn{1}{c}{$E_{\texttt{XC-LYP}}$} \\
\hline
0 & --& $-$2.86164  & $-$2.90374 & $-$2.90651 \\
& & $-$2.86168$^a$ & $-$2.9037$^{b,c}$ &  $-$2.90372$^{d}$\\
\hline
5 & 0.01    & 7.13826  & 7.09616 &  7.09339 \\
 & 0.05    & 7.12818  & 7.08596  & 7.08324 \\
& 0.1    & 7.06951  & 7.02664  & 7.02423 \\
& 0.5    &  4.08392  & 4.02488 & 4.03309 \\
& 1     & 0.82538  & 0.76685  & 0.77690  \\
& 5     & $-$2.48937 & $-$2.53562 & $-$2.53516 \\
& 10    & $-$2.75323 & $-$2.79686 & $-$2.79851 \\
& 50    & $-$2.85693 & $-$2.89911 & $-$2.90182 \\
& 100   & $-$2.86046 & $-$2.90258 & $-$2.90534 \\    \hline

10 & 0.01 & 17.13816 & 17.09606  & 17.09329 \\
 & 0.05 & 17.11773 & 17.07539 & 17.07273 \\
 & 0.1  & 16.99272 & 16.94902  & 16.94701 \\
 & 0.5  & 9.54836  & 9.47907  & 9.49591  \\
 & 1     & 3.39598  & 3.33085  & 3.34730  \\
 & 5     & $-$2.18759 & $-$2.23616 & $-$2.23368 \\
 & 10    & $-$2.65572 & $-$2.70046 & $-$2.70124 \\
 & 50    & $-$2.85224 & $-$2.89450 & $-$2.89716 \\
 & 100   & $-$2.85928 & $-$2.90142 & $-$2.90416\\
\hline

25 & 0.01 & 47.13788 & 47.09577 & 47.09300 \\
 & 0.05 & 47.08465 & 47.04194 & 47.03943  \\
 & 0.1  & 46.70123 & 46.65449 & 46.65398 \\
 & 0.5  & 21.74351  & 21.65988 & 21.69110 \\
   &1   & 9.04943  & 8.97469  & 9.00099   \\
   &    &  &   &  9.00570$^b$\\
&5   & $-$1.45190 & $-$1.50471 & $-$1.49832  \\
 &    &  & & $-$1.49346$^b$\\

&10  & $-$2.39905 & $-$2.44614 & $-$2.44493  \\
&  &  & & $-$2.44032$^b$\\
&50  & $-$2.83840 & $-$2.88088 & $-$2.88339 \\
 &   & & & $-$2.88033$^b$\\
&100 & $-$2.85575 & $-$2.89795 & $-$2.90065 \\ %\hline
 &    &  & & $-$2.89779$^b$\\
\hline\hline
\end{tabular}
\vspace{2mm}
{\footnotesize \\ 
$^{a}$Yakar \emph{et al} \cite{yusuf};
$^{b}$Laughlin and Chu \cite{laughlin09};
$^{c}$Flores-Riveros and Rodriguez-Contreras \cite{2flores};
$^{d}$Bhattacharyya \emph{et al} \cite{bhattacharyya}.
}
\label{tab:1}
\end{table}

%%%%%%%%%%%%%%%%%%%%%%%%%%%%%%%%%%%%%%%%%%%%%%%%%%%%%%%%%%%%%%%%%%%%%%%%%%%%%%%%%%%%%%%%%%%%%%%%%%%%%
%\onecolumngrid
Now to solve the KS equation self-consistently, a highly successful GPS method is employed, that enables an accurate and efficient solution through a non-uniform, optimally distributed spatial grid. 
It is based on approximating an \textit{exact} function $f(x)\in[-1,1]$ by an $N$-th order polynomial $f_{N}(x)$ as,
\begin{equation} \label{eqn:8}
f(x) \cong f_{N}(x)=\sum_{j=0}^{N} f(x_{j})g_{j}(x)\,,
\end{equation}     
in a way such that,
\begin{equation}
\label{eq:9}
f_{N}(x_{j})=f(x_{j})\,. 
\end{equation}
It also guarantees that the approximation is \emph{exact} at the corresponding collocation points. 
Owing to the impenetrable nature of spherical cavity of radius $r_c$, we expand the radial region within the range $[0, r_c]$. 
Therefore, to facilitate the numerical treatment, we have mapped $r_c~(r \in [0, r_c])$ onto the interval $[-1,1]$ using a suitable nonlinear mapping function,  
\begin{equation}
\label{eq:10}
r=r(x)=L\ \ \frac{1+x}{1-x+\alpha},
\end{equation}
where, $L$ and $\alpha=2L/r_{c}$ are two mapping parameters.

In Legendre pseudo-spectral method, which is the one used here, $x_0$ and $x_N$ take values of $-1$ and 1 respectively, while all other roots $x_{j}(j=1,....,N-1)$ are determined from the first-order derivative of Legendre polynomial $P_{N}(x)$ following the equation,
\begin{equation}
\label{eq:11}
P_{N}'(x_{j})=0.
\end{equation}
The cardinal functions, $g_{j}(x)$, are evaluated from,
\begin{equation}
\label{eq:12}
g_{j}(x)=-\frac{1}{N(N+1)P_{N}(x_{j})}\frac{(1-x^{2})P_{N}'(x)}{(x-x_{j})},
\end{equation} 
which satisfies the condition,
\begin{eqnarray}
\label{eq:13}
    g_{j}(x_{j'})=\delta_{j',j}. 
\end{eqnarray}
This procedure leads to a symmetric eigenvalue problem that is solved using standard numerical algorithms to accurately determine the eigenvalues and eigenfunctions.
For further methodological details, we refer the reader to the articles \cite{roy2004b,roy2004a,roy2014}, and references therein.  
The impenetrable nature of spherical cavity in the system is achieved by ensuring the total electron density ($n_i$ denotes the occupation number of $i$th orbital), 
\begin{eqnarray}
\label{eq:14}
    \rho(\vec{r}) = \sum_{i=1}^{N} n_{i}|\psi_{i}(\vec{r})|^2, 
\end{eqnarray}
dissipating at the cavity boundary, satisfying a Dirichlet boundary condition, $\psi_{nl}(0) = \psi_{nl}(r_c) = 0$. 
In the above equation, $n_{i}$ represents the occupation number of \textit{i}th orbital. 
It follows the normalization criterion,
\begin{eqnarray}
\label{eq:15}
    \int \rho(\vec{r}) d\vec{r} = N. 
\end{eqnarray}
All calculations in this work are performed numerically, with convergence in energy and potential ensured by systematically varying grid parameters, total number of radial points and the maximum radial range. 

\section{Results and discussion}\label{sec:results}
Before going to a detailed discussion, a few general remarks may be made at the outset. A simple convergence criteria of $10^{-7}$ in total energy and $10^{-6}$ in 
potential is adopted during the self-consistent iterative procedure. The GPS parameters employed are as follows: number of radial grid points, $n_{r}=300$ and $L$=1;
these are maintained uniformly throughout the entire confinement region. All the data are checked rigorously through a convergence test. 
Besides these, potential parameters, such as cavity width ($R \in [0.001,100]$) and cavity depth ($V_0$) are also chosen judiciously, so as to include the asymptotic 
limit of $R$. Calculations are restricted to two different $V_0$ (5 and 10), which suffice to bring out the essential physics of the system. In terms of XC effects, 
three sets of results are reported, namely (i) exchange-only (XX) (ii) XC including Wigner functional (XC-WIG) and (iii) XC with LYP functional (XC-LYP). 

%%%It should be mentioned that a handful literature is available for the ground state energies of free He-like systems ($Z=2-18$) \cite{mondal24, pest07, saha2002, thakkar77, utpal99}. 

Next, to establish the consistency of our method and for ease of presentation, at first, estimated ground-state energies for H, He-like ions ($Z=2-18$) as well as Li and Be
atom are offered. The analysis is then extended for $1s2s$ ($^{1,3}$S) states of He. Our current results on H atom corroborate very closely to those reported in articles 
\cite{zhou25,liu24} and not presented here further. The energy analysis of two-electron atoms in GP has been made rather scarcely \cite{xie08,laughlin09}.
However, results could not be found for information-theoretic measures under GP confinement, for which we 
%%%%\cite{kdsen,garza,gadre,martinez,chakaldar}. (check the references ???)
make a comprehensive study in He-like ($Z=2-18$) systems. Among them, for future purposes, we will make the forthcoming analysis by restricting to two 
representative species of the series, \emph{viz.,} He ($Z=2$) and O$^{6+}$ ($Z=8$). The energies of Li, Be, as well as excited states of He $1s2s$ ($^{1,3}$S) 
under GP have been investigated for the first time. All results are given in atomic units, unless stated otherwise. 

\setlength{\tabcolsep}{8pt}
\begin{table*}[!th]
\centering
\caption{Ground-state energy of Li and Be in GP, for different $V_{0}$ and $R$. All quantities are in a.u.}
\begin{tabular}{ccrrrrrrrr}
%\toprule
\hline\hline
 & &  & \multicolumn{3}{c}{Li} && \multicolumn{3}{c}{Be}\\ 
\cline{4-6}\cline{8-10}
$V_{0}$ &$R$ && \multicolumn{1}{c}{$E_{\texttt{XX}}$}  &\multicolumn{1}{c}{$E_{\texttt{XC-WIG}}$} & \multicolumn{1}{c}{$E_{\texttt{XC-LYP}}$} && \multicolumn{1}{c}{$E_{\texttt{XX}}$}  &\multicolumn{1}{c}{$E_{\texttt{XC-WIG}}$} & \multicolumn{1}{c}{$E_{\texttt{XC-LYP}}$} \\

\hline
5  & 0.01 && 7.568185  & 7.503620  & 7.501997  && 5.427818   & 5.336362   & 5.328432   \\
  & 0.05 && 7.533386  & 7.468625  & 7.467071  && 5.347621   & 5.255883   & 5.248059   \\
  & 0.1  && 7.353286  & 7.287545  & 7.286390  && 4.977645   & 4.884684   & 4.877379   \\
  & 0.2  && 6.489203  & 6.419506  & 6.420207  && 3.540483   & 3.443817   & 3.438359   \\
  & 0.3  && 5.229945  & 5.156369  & 5.159292  && 1.892005   & 1.792869   & 1.788919   \\
  & 0.5  && 2.880739  & 2.804360  & 2.809295  && $-$0.588826  & $-$0.689455  & $-$0.693719  \\
  & 0.7  && 1.291470  & 1.214910  & 1.217614  && $-$2.180804  & $-$2.284497  & $-$2.291917  \\
  & 1    && $-$0.302261 & $-$0.382425 & $-$0.385324 && $-$4.260143  & $-$4.372430  & $-$4.379369  \\
  & 2    && $-$3.681165 & $-$3.761288 & $-$3.764059 && $-$9.205826  & $-$9.318368  & $-$9.321962  \\
  & 3    && $-$5.108922 & $-$5.185746 & $-$5.189728 && $-$11.296037 & $-$11.404332 & $-$11.409397 \\
  & 5    && $-$6.233635 & $-$6.306614 & $-$6.311887 && $-$12.933212 & $-$13.036228 & $-$13.043383 \\
  & 7    && $-$6.677927 & $-$6.748867 & $-$6.754481 && $-$13.570644 & $-$13.670715 & $-$13.678818 \\
  & 10   && $-$6.980295 & $-$7.049506 & $-$7.055030 && $-$13.995481 & $-$14.092989 & $-$14.101643 \\
  & 50   && $-$7.398898 & $-$7.464194 & $-$7.466726 && $-$14.538182 & $-$14.630273 & $-$14.638450 \\
  & 100  && $-$7.422552 & $-$7.487356 & $-$7.489245 && $-$14.562733 & $-$14.654364 & $-$14.662364 \\
\hline
10 & 0.01 && 22.567825 & 22.503258 & 22.501635 && 25.426915  & 25.335457  & 25.327528  \\
 & 0.05 && 22.497273 & 22.432308 & 22.430827 && 25.264362  & 25.172332  & 25.164617  \\
 & 0.1  && 22.115638 & 22.048632 & 22.047996 && 24.486210  & 24.391666  & 24.385037  \\
 & 0.2  && 20.116895 & 20.041759 & 20.045184 && 21.320444  & 21.218710  & 21.215876  \\
 & 0.3  && 17.176541 & 17.095115 & 17.102726 && 17.769838  & 17.664350  & 17.664018  \\
 & 0.5  && 12.234510 & 12.150552 & 12.159555 && 12.766110  & 12.657768  & 12.655079  \\
 & 0.7  && 9.113399  & 9.026559  & 9.028485  && 9.236861   & 9.117575   & 9.112855   \\
 & 1    && 5.392024  & 5.299552  & 5.302148  && 3.988944   & 3.861441   & 3.863487   \\
 & 2    && $-$1.162442 & $-$1.248899 & $-$1.247388 && $-$5.527619  & $-$5.648438  & $-$5.646299  \\
 & 3    && $-$3.587107 & $-$3.668640 & $-$3.669843 && $-$9.079106  & $-$9.193688  & $-$9.194847  \\
 & 5    && $-$5.447243 & $-$5.523431 & $-$5.527423 && $-$11.798157 & $-$11.905641 & $-$11.910615 \\
 & 7    && $-$6.175916 & $-$6.249320 & $-$6.254396 && $-$12.854790 & $-$12.958400 & $-$12.965169 \\
 & 10   && $-$6.671872 & $-$6.742935 & $-$6.748510 && $-$13.564802 & $-$13.665038 & $-$13.673034 \\
 & 50   && $-$7.371118 & $-$7.436869 & $-$7.440010 && $-$14.507224 & $-$14.599822 & $-$14.608173 \\
 & 100  && $-$7.414240 & $-$7.479235 & $-$7.481362 && $-$14.554371 & $-$14.646167 & $-$14.654231\\
\hline\hline
\end{tabular}
\label{tab:2}
\end{table*}

\subsection{Energy analysis}\label{sec:energy}
\begin{figure*} %[!th]
\centering
\begin{tabular}{cccc}
(a)  &  (b) & (c) & (d)\\
\includegraphics[width=0.235\textwidth]{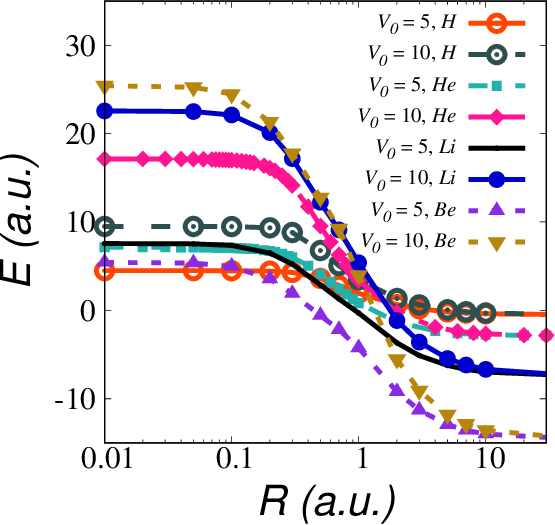}&  
\includegraphics[width=0.21\textwidth]{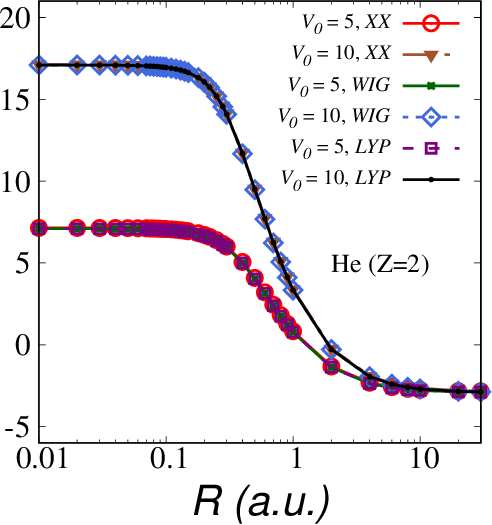}&
\includegraphics[width=0.21\textwidth]{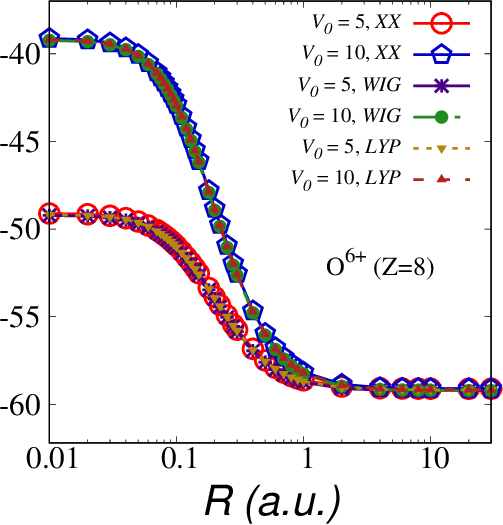}&
\includegraphics[width=0.22\textwidth]{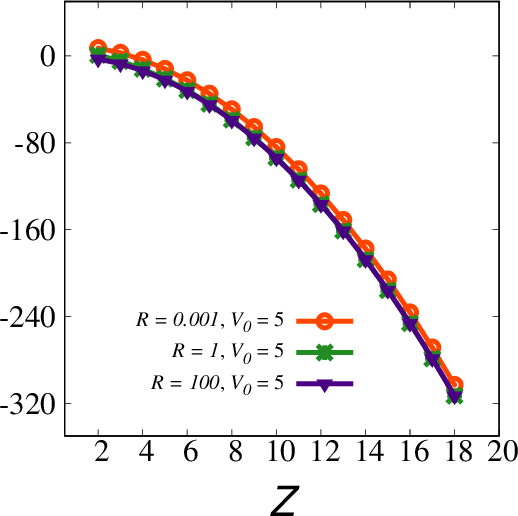}
%\includegraphics[width=0.38\textwidth]{Fig2.eps} 
%\hspace{0.1cm}
%\includegraphics[width=0.35\textwidth]{Fig1d.eps}
\end{tabular}
	\caption{$E$ vs. $R$, for $V_{0}=5, 10$ for ground states of H, He, Li and Be, with XX, is shown in panel (a); for ground states of He and O$^{6+}$ with XX, XC-WIG, XC-LYP in 
	(b) and (c). Panel (d) gives energy plot for He-like ions, with XX, for $V_0=5$.}
%\caption{Variation of $E$ for He, O$^{6+}$, Al$^{11+}$ and Ar$^{16+}$ w.r.t. $R$, for $V_{0}=5$ and $10$, in panels (a)–(d). See text for details.}
\label{fig:1}
\end{figure*}

The ground-state energies of H, He, Li, Be and low-lying $1s2s$ $^{1,3}S$ excited states of He, confined within a GP are computed for different sets of ($V_{0}, R$). 
The nature of variation can be well understood by analyzing the corresponding asymptotic limits at first. It is observed that for larger $R \rightarrow \infty$, 
irrespective of $V_{0}$, energy of a system approaches that of the respective free counterparts. In contrast, as it tends to zero ($R \rightarrow 0$), energy turns 
to positive values, mainly due to a rise in kinetic energy. Thus, energy decreases monotonically with $R$. Evidently, as $R \rightarrow 0$, the confinement 
potential $v_{\text{conf}} \rightarrow V_0$. This effectively increases energy by an amount equivalent to $2V_{0}$. On the other hand, as $R \rightarrow \infty$, 
the potential $v_{\text{conf}} \rightarrow 0$, which corresponds to the energy of an unconfined system. For H atom, GPS energies deliver excellent agreement with 
previously reported result \cite{zhou25,liu24,ma25} and hence omitted here to save space. Thus, our primary focus is on multi-electron systems, namely, He, Li, Be. 
The calculated ground-state energy for He has been listed in Table~\ref{tab:1}, whereas that for Li and Be in Table~\ref{tab:2}. In first table, three $V_0$ 
(5, 10, 25), while in second table, two (5, 10) $V_0$'s are selected. For He, some reference results are available, both at HF and correlated level, which are quoted 
appropriately, but for Li, Be, to the best of our knowledge, no results have been reported as yet. The agreement between the current and reference energies 
are, in general, quite good. It is evident that,
for a given $V_0$, energy transition from negative to positive occurs at $R\approx 5$ for He, whereas for Li and Be it occurs much earlier, around $R\approx 1$. 
Beyond $R\approx 50$, the system starts to behave as a free system irrespective of $V_{0}$. Now these are graphically depicted in Fig.~\ref{fig:1}. In panel (a), the 
XX energy of ground state for H, He, Li and Be, w.r.t. $R$ and $V_{0}$ are plotted; correlated results are omitted as they do not reflect much qualitative 
change in the pattern behavior. 
Note that the $R$ axis in this and all other figures throughout the article, is plotted in logarithmic scale, for convenience.  
For a smaller $R$ ($ \lessapprox 0.1$), individual energies remain quite distinctly separated; however there is significant overlap amongst them for the 
region $0.25<R<5$. 
Finally for $R \gtrsim 10$, plots for individual atoms tend to merge and remain practically unaltered thereafter. Now to get a clearer picture, panel (b) depicts the plots for 
He in case of $V_0=5$ and 10 for XX, XC-WIG and XC-LYP calculations. It is noticed that the plots for two $V_0$'s make two separate families, i.e., the three energies 
for a given $V_0$ remain quite close to each other. And predictably, energies reach that of the corresponding free system for a sufficiently large $R$. 
The effect of confinement is more prominent between cavity widths $R \sim 0.1$ to $R \sim 10$, providing an effective width range $\Delta=R_2-R_1\sim 9.9$. 
When $Z$ increases, as illustrated for O$^{6+}$ in panel (c), ($-\frac{Z}{r}$) potential becomes stronger, which pulls the electrons closer to impurity center of 
GP and hence, the bound-state energy rises, as expected. On the other hand, kinetic energy increases and effective potential interaction energy decrease due to a 
stronger localization. The pattern is similar for all systems under consideration, but the effective range varies as $Z$ increases. 
For He-iso-electronic series, both $R_1$ and $R_2$ shift towards the left; so the effective width $\Delta$ decreases with increment in $Z$. 

In order to get an explicit dependence on $Z$, panel (d) of Fig.~\ref{fig:1} displays energy (XC-WIG) changes in He-like ions. For $Z=2-18$, three $R$ values 
(0.001, 1, 100) are provided for a fixed $V_0=5$. Evidently, for low-$Z$ ($Z<8$), both $V_{0}$ and $R$ significantly modify the charge distribution.  
In contrast, for high-$Z$ ($Z>8$) systems, the density is already extremely compact due to a stronger ($-\frac{Z}{r}$) interaction. 
Therefore, the characteristic size of the electron cloud shrinks and it becomes much smaller than the Gaussian width for most cases.
Hence, the confinement impacts rather mildly on the system, even if $V_{0}$ is large. Thus, the influence of GP diminishes with increasing $Z$ and ($-\frac{Z}{r}$) 
dominates over both the external GP and electron-electron repulsion. This leads to a rapid decrease in total energy (more negative), which reflects a stronger binding 
because electrons are now pulled tightly toward the nucleus.This reduces correlation energy in all cases. For a better understanding, energies of He-iso-electronic series 
are fitted as a function of $Z$, offering the following parabolic equation: 
\begin{equation} \label{fit:evsz}
    E = E_0 - E_1 Z^2. 
\end{equation}
The calculated fitting parameters $E_{0}$ and $E_1$ are given at the top of Table~\ref{tab:3} along with the reduced $\chi^2$ values. The latter is close to 1 for all $R$, 
showing the goodness in fitting. Moreover, the offset $E_0$ monotonically decreases, whereas $E_1$ almost remains invariant.
%%The detailed variation of the present data (symbols) and the fitted data (solid lines) is clearly depicted in Fig.~\ref{fig:EvsZ}. 

\begin{figure}[t]
    \centering
    \begin{tabular}{c}
\includegraphics[width=0.35\textwidth]{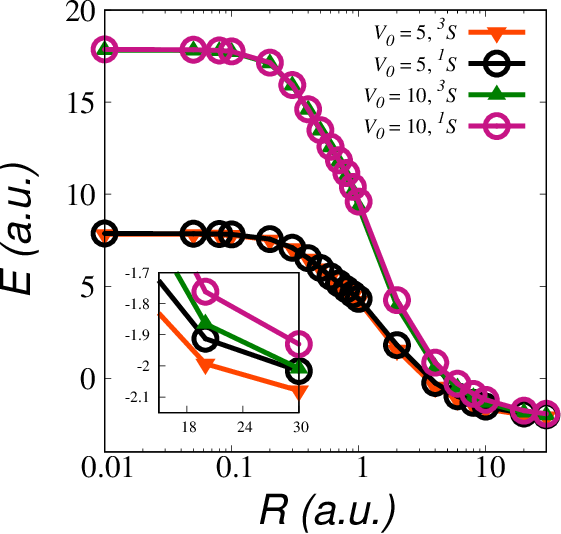}
    \end{tabular}
	\caption{He 1s2s ($^1S$, $^3S$) energies vs. $R$ for $V_0=5$ and 10, in GP, at the XX level of theory. See text for details.}
    \label{fig:EvsZ} 
\end{figure}

After a discussion of ground state, we now extend the analysis to excited state, focusing on low-lying $1s2s$ ($^{1,3}$S) of He atom. 
The energy changes with confinement parameters is tabulated in Table~\ref{tab:4}. Once again a large range of $R$ is covered for same two $V_0$ of previous table. No 
reference results could be found in the literature for direct comparison. The trend is shown in Fig.~\ref{fig:EvsZ} for XX energies, which is quite similar to 
that found for ground state. In accordance with Hund's rule, the triplet state remains lower in energy than the singlet. 

\setlength{\tabcolsep}{3pt}
\begin{table}[b]
\centering
\caption{Fitting parameters for $E$ vs $Z$ (Eq.~(\ref{fit:evsz}), $S_{r}$ vs $Z$ (Eq.~\eqref{srvsz}), $S_{\texttt{r}_{\texttt{corr}}}$ vs $Z$ (Eq.~\eqref{srcorrvsz}), $S_{p}$ vs $Z$ (Eq.~\eqref{spvsz}), $S_{\texttt{p}_{\texttt{corr}}}$ vs $Z$ (Eq.~\eqref{spcorrvsz}) and  $I_{r}$ vs $Z$ (Eq.~\eqref{irvsz}).} 
\begin{tabular}{cccccc}
\hline
\hline
$R$ &$E_{0}$ & $E_1$ & --- & --- &$\chi^2$ \\
\hline
0.001 & 12.361764 & 0.970421 & --- & --- & 0.999957   \\
1     & 3.901002  & 0.976787 & --- & --- & 0.999991   \\
100   & 2.362071  & 0.970423 & --- & --- & 0.999957  \\
%\toprule
%\cline{2-6}
\hline
$R$&$s_{r_0}$ & $s_{r_1}$ & $Z_1$ & --- &$\chi^2$ \\
\hline
%\multicolumn{5}{c}{$r-$ space}\\
%\cline{2-5}
0.001 & $-$4.929970 & 9.914077 & 6.575855 & --- &0.996062 \\
1     & $-$5.840566 & 8.643767 & 9.959466 & --- &0.999847 \\
100   & $-$4.930956 & 9.911910 & 6.579175 & --- &0.996076 \\
%\cline{2-6}
%\multicolumn{5}{c}{$r-$ space}\\
%\cline{2-5}
\hline
$R$&$s_{p_0}$ & $s_{p_1}$ &$Z_2$&--- &$\chi^2$ \\
%\cline{2-5}
\hline
0.001 & 1.646580 & 9.863257 & 6.614804 &--- & 0.996148 \\
1     & 3.670430 & 8.674153 & 9.706084 &--- & 0.999801 \\
100   & 1.649101 & 9.861534 & 6.617502 &--- &0.996159 \\
\hline
 $R$&$s^{\text{r,corr}}_0$ & $s^{\text{r,corr}}_1$ & $Z_3$ & --- &$\chi^2$ \\
\hline
%\multicolumn{5}{c}{$r-$ space}\\
%\cline{2-5}
0.001 & 0.000682 & 0.157745 & 1.243525 & --- & 0.995404 \\
1     & 0.000152 & 0.015349 & 3.145199 & --- &0.999854 \\
100   & 0.000679 & 0.156096 & 1.248370 & --- &0.995448 \\
%\cline{2-6}
\hline
$R$&$s^{\text{p,corr}}_0$ & $s^{\text{p,corr}}_1$ &$Z_4$& --- & $\chi^2$ \\
%\cline{2-6}
\hline
0.001 & 0.000500 & 0.100980 & 1.419568 & --- & 0.997425 \\
1     & 0.000136 & 0.014667 & 3.047317 & --- & 0.999831 \\
100   & 0.000502 & 0.101360 & 1.417101 & --- & 0.997396 \\
%\cline{2-6}
\hline
$R$&$I_{r_0}$ & $I_{r_1}$ & --- & --- &$\chi^2$ \\
\hline
0.001 & $-$9.460890 & 3.883442 &  --- & --- &0.999956 \\
1     & $-$1.526871 & 3.852985 &  --- & --- &0.999837 \\
100   & $-$9.458309 & 3.883427 &  --- & --- &0.999956 \\
%\cline{2-6}
\hline
 $R$&$I^{\text{r}_\text{corr}}_{0}$ & $I^{\text{r}_\text{corr}}_{1}$ & $I^{\text{r}_\text{corr}}_{2}$  & $I^{\text{r}_\text{corr}}_{3}$ & $\chi^2$ \\
%\cline{2-6}
\hline
0.001 & 0.149181 & 0.024600 & $-$0.005126 & 0.000331  & 0.996532 \\
1     & 0.042709 & 0.049934 & $-$0.007213 & 0.000397  & 0.998760  \\
100   & 0.148384 & 0.024962 & $-$0.005182 & 0.000335  & 0.996430  \\
\hline
\hline
\end{tabular}
%%%\label{tab:4}
\label{tab:3}
\end{table}

\begin{figure*}[!th]
    \centering
    \begin{tabular}{ccc}
       (a) & (b) & (c) \\
\includegraphics[width=0.31\textwidth]{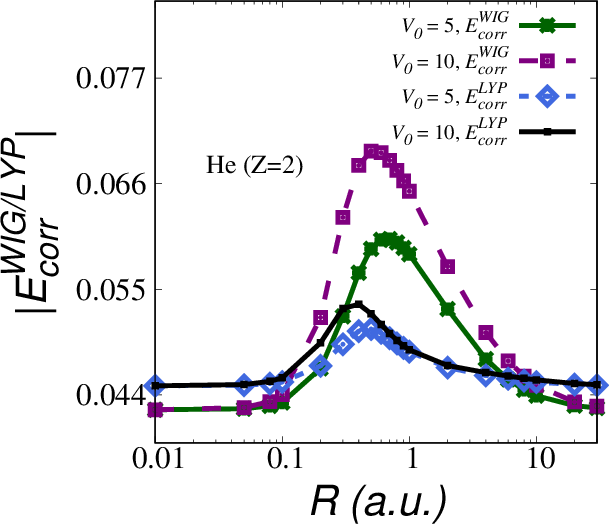}& \hspace{-0.3cm}
\includegraphics[width=0.27\textwidth]{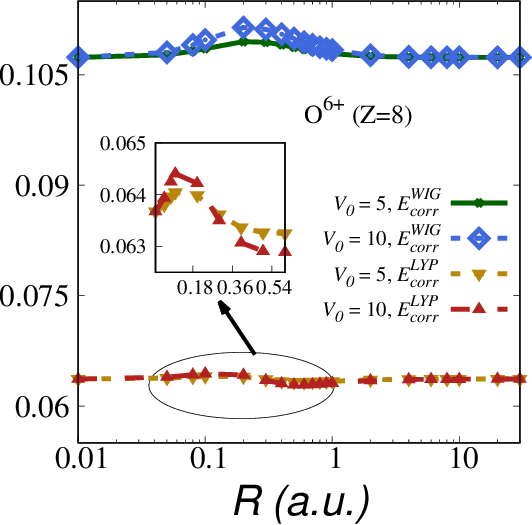}& \hspace{-0.3cm}
\includegraphics[width=0.31\textwidth]{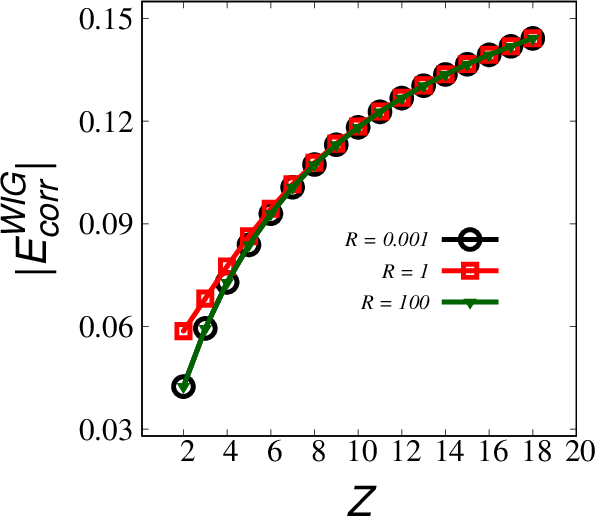}
%(c) & (d)\\
%\includegraphics[width=0.35\textwidth]{Fig3d.eps}
    \end{tabular}
	\caption{The behavior of $|E^{WIG/LYP}_{\texttt{corr}}|$ w.r.t. $R$, for $V_{0}=5$ and $10$. Panels (a), (b) offer these for He and O$^{6+}$. Panel (c) shows 
	$|E^{WIG}_{\texttt{corr}}|$ w.r.t. $Z$ with $V_{0}=5$ in the He-iso-electronic series. See text for details.}
    \label{fig:3}
\end{figure*}

%%%%%Next, to estimate correlation energy contributions to total energy, the quantity $E_{\texttt{corr}}$ is defined as the difference between our calculated correlated 
%%%%%total energy (XC-WIG/XC-LYP) and exchange-only (XX) energy i.e.,
%%%%%%%%\begin{equation}
%%%%% $E_{\texttt{corr}} \approx E_{\texttt{XC}} - E_{\texttt{XX}}$. Note that, E$_{\texttt{XX}}$ from this work-function potential provides near-HF quality results, 
%%%%%which has been confirmed in a large number of previous works \cite{roy97,roy97a,mondal}. Thus, this definition can be adopted without any loss of generality. %%%It is observed that the error in $E_{\texttt{corr}}$ estimated here lies below 1\% of the total correlation if calculated with the true Hartree-Fock energy. 
%%%More importantly, loss of this amount of $E_{\texttt{corr}}$ does not contribute much to the total variation of the correlation measure and other physical properties evaluated here.
Next we proceed to understand the nature of electron correlation, obtained from the two functionals. Towards this goal, the estimated correlated energies, 
	$|E^{WIG/LYP}_\texttt{corr}|$ w.r.t. $V_{0}$ and $R$ are shown, for He and O$^{6+}$ in panels (a), (b) of Fig.~\ref{fig:3}. 
Apparently, the correlation effect seems to influence the intermediate region (characterized by a maximum) much more of the underlying confinement. The $R$-variation of 
$|E_\texttt{corr}|$ for a fixed $V_0$ shows quite similarity for two correlation functionals. Generally, the two give distinctly separate results, with 
XC-WIG remaining at the top of XC-LYP. Moreover, for a given correlation functional, the higher maximum corresponds to larger $V_{0}$, because in the latter scenario, the 
system gets more localized. %As $Z$ increases, the maxima in the intermediate $R$ flattens, so much so that, for Ar$^{16+}$, the plots virtually become like straight line. 
This is clearly visible from (a), where the peaks for XC-WIG are observed at $R\sim 0.6, 0.5$ for $V_{0}=5, 10$, whereas the same for XC-LYP can be seen at 
$R=0.5$ and 0.4 for same $V_0$. Interestingly, as one goes to lower $R$, the two results seem to cross each other, which is perhaps due to the nature of two 
functionals. In the higher side of $R$, however, GP becomes negligible, and the system approaches free limit. Thus $E_{\texttt{corr}}$ is 
governed mainly by the atomic density profile and its long-range tail. %%%Hence, the two functionals approaches the free-atom $E_{\texttt{corr}}$ with different asymptotic behavior, which leads to persistent crossings between their curves even at large confinement widths.

Moving to higher impurity $Z$, representative results are portrayed for O$^{6+}$ in panel (b). 
	Here, an interesting feature is observed: unlike the case of He, no crossing occurs between $|E^{WIG}_\texttt{corr}|$ and $|E^{LYP}_\texttt{corr}|$ for a given $V_0$. 
This is because by increasing the impurity $Z$, ($-\frac{Z}{r}$) strongly localizes the electrons near the center. 
In this case, confinement affects the density only over a narrow range of $R$, outside of which, it changes very little with width, and $E_{\texttt{corr}}$ varies 
smoothly. %%%Differences between functionals become negligible because of the same sharply peaked density. 
Like He atom, $|E^{WIG}_\texttt{corr}|$ records a maximum for both $V_{0}$. However, some sensitivity of $V_{0}$ persists, as a different nature can be observed for 
$|E^{LYP}_\texttt{corr}|$ in the inset plot. Furthermore, the maximum shifts toward the left, and the strength of electron correlation due to compactness of 
density near nucleus increases. 
%%Consequently, $E_{\text{corr}}$ increases with $Z$, even though its relative contribution to the total energy becomes smaller as has been depicted in Fig.~\ref{fit:ecorrvsz}.
%%%Now it is very interesting to discuss the effect of $E_{\texttt{corr}}$ on He-like ($Z=2-18$) systems. 
Next, the qualitative behavior of $E_{\texttt{corr}}$ as function of $Z$ for $R$ ($R=0.001,1$ and 100) is depicted in panel (c). 
However, unlike low $Z$, in case of larger $Z$, both electrons in the systems are forced into an increasingly compact region near nucleus, which enhances their mutual 
repulsion and makes correlated motion essential to lower overall energy. 

\setlength{\tabcolsep}{8pt}
\begin{table*}[!th]
\centering
\caption{Energy of He 1s2s $^{1,3}$S states in GP, for different $V_{0}$ and $R$. All quantities are in a.u.}
%%%For the comparison purposes, we have included the following references, $a$: Ludena \cite {ludena}, $b$: Claude et al. \cite{lesech}, $c$: Gimarc \cite{gimarc}, $d$: Waugh et al. \cite{waugh}, $e$: Laughlin \cite{laughlin09}, $f$: Flores-Riveros \cite{2flores}, and $g$: Bhattacharyya \cite{bhattacharyya}.} 
\begin{tabular}{ccrrrrrrrrr}
\hline\hline
 & && \multicolumn{2}{c}{$E_{\texttt{XX}}$}  ~ &&\multicolumn{2}{c}{$E_{\texttt{XC-WIG}}$} ~ && \multicolumn{2}{c}{$E_{\texttt{XC-LYP}}$} ~\\ 
 \cline{4-5}\cline{7-8}\cline{10-11}
%\hline
$V_{0}$ &$R$ && \multicolumn{1}{c}{$^3${S}} & \multicolumn{1}{c}{$^1${S}} && \multicolumn{1}{c}{$^3${S}} & \multicolumn{1}{c}{$^1${S}} && \multicolumn{1}{c}{$^3${S}} & \multicolumn{1}{c}{$^1${S}}\\
\hline
5  & 0.1 && 7.776346  & 7.815352  && 7.750576  & 7.791238  && 7.750331  & 7.797931  \\
  & 0.5 && 5.965658  & 5.991956  && 5.932998  & 5.960472  && 5.936273  & 5.966352  \\
  & 0.1 && 7.776346  & 7.815352  && 7.750576  & 7.791238  && 7.750331  & 7.797931  \\
  & 0.3 && 7.056716  & 7.092836  && 7.026541  & 7.064163  && 7.028138  & 7.070414  \\
  & 0.5 && 5.965658  & 5.991956  && 5.932998  & 5.960472  && 5.936273  & 5.966352  \\
  & 0.7 && 5.121453  & 5.153713  && 5.088302  & 5.122151  && 5.090552  & 5.126531  \\
  & 1   && 4.198992  & 4.313241  && 4.162656  & 4.280044  && 4.160411  & 4.277819  \\
  & 2   && 1.546074  & 1.798471  && 1.506834  & 1.759858  && 1.506473  & 1.759205  \\
  & 4   && $-$0.469289 & $-$0.255303 && $-$0.504375 & $-$0.290180 && $-$0.506593 & $-$0.292180 \\
  & 6   && $-$1.156251 & $-$0.979911 && $-$1.188929 & $-$1.012467 && $-$1.192240 & $-$1.015001 \\
  & 8   && $-$1.482902 & $-$1.333848 && $-$1.514097 & $-$1.364897 && $-$1.517874 & $-$1.367290 \\
  & 10  && $-$1.667905 & $-$1.538795 && $-$1.698098 & $-$1.568771 && $-$1.702026 & $-$1.570706 \\
  & 20  && $-$1.993469 & $-$1.912630 && $-$2.021321 & $-$1.939860 && $-$2.024581 & $-$1.938670 \\
  & 50  && $-$2.134362 & $-$2.084195 && $-$2.160535 & $-$2.109197 && $-$2.161820 & $-$2.103597 \\
  & 100 && $-$2.162522 & $-$2.120377 && $-$2.188144 & $-$2.144556 && $-$2.188741 & $-$2.137943 \\ \hline
10 & 0.1 && 17.720432 & 17.759921 && 17.694185 & 17.735351 && 17.694061 & 17.742017 \\
 & 0.5 && 13.459561 & 13.482343 && 13.422281 & 13.446151 && 13.428159 & 13.453303 \\
 & 0.1 && 17.720432 & 17.759921 && 17.694185 & 17.735351 && 17.694061 & 17.742017 \\
 & 0.3 && 15.903331 & 15.932908 && 15.868499 & 15.899316 && 15.872692 & 15.906195 \\
 & 0.5 && 13.459561 & 13.482343 && 13.422281 & 13.446151 && 13.428159 & 13.453303 \\
 & 0.7 && 11.764873 & 11.843913 && 11.725268 & 11.807286 && 11.726225 & 11.808260 \\
 & 1   && 9.323181  & 9.601365  && 9.276293  & 9.555880  && 9.278956  & 9.557730  \\
 & 2   && 3.918499  & 4.243531  && 3.874161  & 4.199620  && 3.878185  & 4.203328  \\
 & 4   && 0.584587  & 0.849165  && 0.546215  & 0.810988  && 0.546285  & 0.810988  \\
 & 6   && $-$0.515962 & $-$0.295157 && $-$0.551193 & $-$0.330304 && $-$0.553135 & $-$0.332006 \\
 & 8   && $-$1.038665 & $-$0.849450 && $-$1.071973 & $-$0.882710 && $-$1.074939 & $-$0.885058 \\
 & 10  && $-$1.335759 & $-$1.170327 && $-$1.367764 & $-$1.202268 && $-$1.371268 & $-$1.204739 \\
 & 20  && $-$1.864627 & $-$1.761446 && $-$1.893563 & $-$1.790011 && $-$1.897366 & $-$1.790706 \\
 & 50  && $-$2.101956 & $-$2.043663 && $-$2.128618 & $-$2.069347 && $-$2.130587 & $-$2.065065 \\
 & 100 && $-$2.152526 & $-$2.107371 && $-$2.178367 & $-$2.131887 && $-$2.179206 & $-$2.125569\\
\hline\hline
\end{tabular}
%\label{tab:3}
\label{tab:4}
\end{table*}

\subsection{Shannon information entropy} \label{sec:Shannon}
In $r$-space, Shannon entropy measures the spread or spatial delocalization of an electronic cloud in a quantum system, or simply provides the measure of lack of 
	structure in it. The explicit knowledge of $r$-space density $\rho(\vec{r})$ from Eq.~\eqref{eq:14} is used to quantify $S_r$ as,
\begin{equation}
\label{eq:17}
S_r =  -\int_{{\mathcal{R}}^3} \rho(\vec{r}) \ln \rho(\vec{r}) d \vec{r}.
\end{equation}
The $p$-space wave function, $\phi_i(\vec{p})$, likewise, is obtained by Fourier transforming the $r$-space counterpart, as follows:
\begin{equation}
\label{eq:18}
\phi_i(\vec{p}) =  \int_{{\mathcal{R}}^3} \psi_i(\vec{r}) \ e^{i\vec{p} .\vec{r}} \mathrm{d} \vec{r}.
\end{equation}
The corresponding Shannon entropy $S_{p}$, in $p$-space is:
\begin{equation}
\label{eq:25}
S_{p} = -\int_{{\mathcal{R}}^3} \Pi(\vec{p}) \ \ln [\Pi(\vec{p})] \ \mathrm{d} \vec{p},
\end{equation}
where $\Pi(\vec{p})$ signifies the $p$-space density. Sum of Eqs.~\eqref{eq:17} and \eqref{eq:25}, yields Shannon entropy sum $S_{t}$ as,
\begin{equation}
\label{eq:26}
S_{{t}} = S_{r}+S_{p}. %\geq d(1+\ln\pi),
\end{equation}
This can be used for verification of famous Bialynicki-Birula-Mycielski (BBM) inequality i.e., $S_t\geq 3(1+ļn ~\pi)$ {\cite{bial}}. It gives an equivalent uncertainty principle based on Shannon 
information entropy. Lower uncertainty corresponds to a smaller $S_{t}$, which reflects a more localized spatial charge distribution. This leads to a higher probability of 
accurately predicting the localization of electron. 
%%%We have tested and verified the BBM inequality and observed the inequality to remain valid for all systems and all potential parameters chosen herein.

\begin{figure*}[!th]
\centering
\begin{tabular}{cccc}
 (a)  &  (b) &  (c) & (d)\\ 
\includegraphics[width=0.24\textwidth]{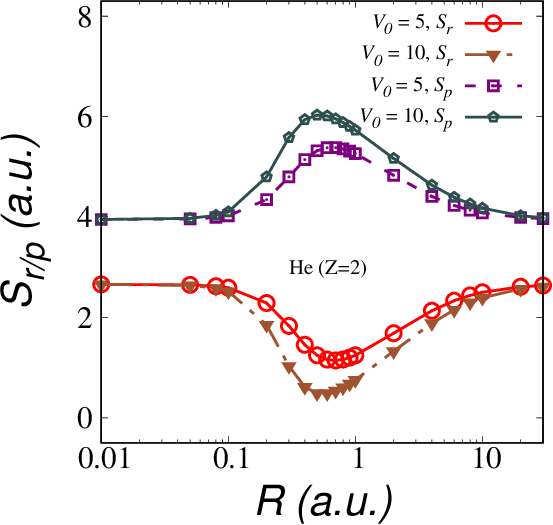}& \hspace{-0.5cm}
\includegraphics[width=0.21\textwidth]{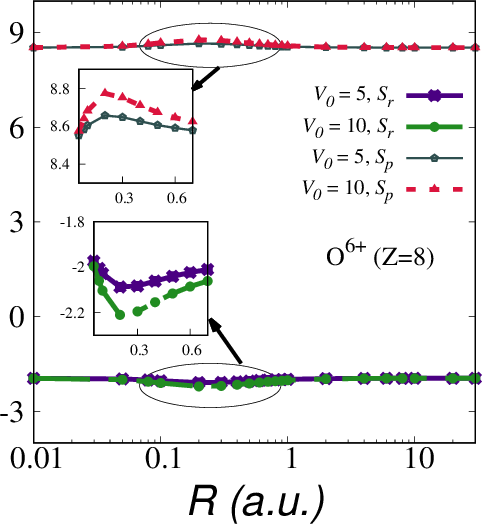}&\hspace{-0.5cm}
\includegraphics[width=0.24\textwidth]{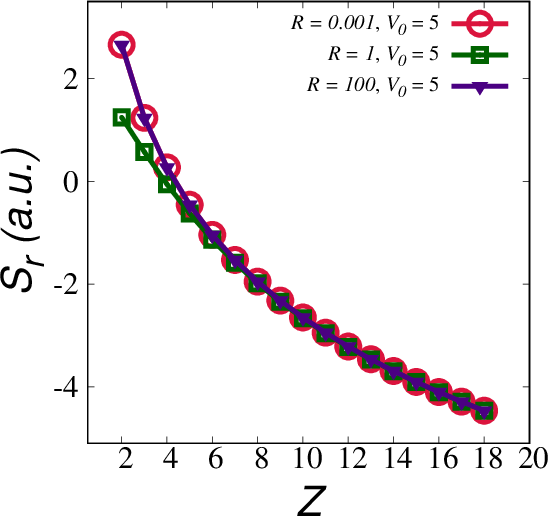}& \hspace{-0.5cm}
\includegraphics[width=0.25\textwidth]{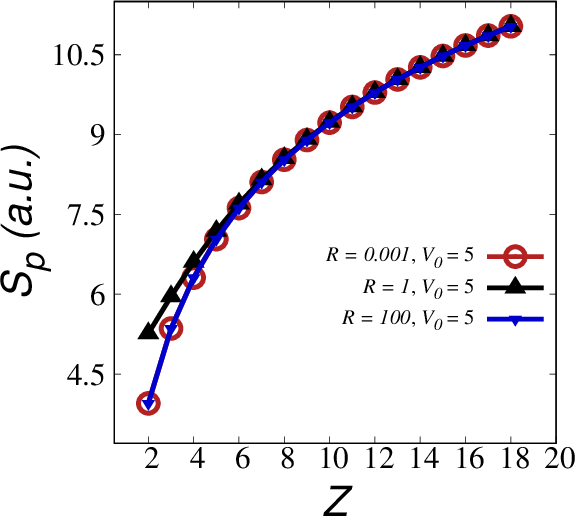}
\end{tabular}
	\caption{Variation of $S_{\texttt{r/p}}$ w.r.t. $R$ with XC-WIG, for $V_{0}=5$ and $10$. Panels (a) and (b) refer to He and O$^{6+}$. Panel (c) and (d) show $S_{r}$ w.r.t. $Z$, and 
	$S_p$ w.r.t. $Z$, for XC-WIG. In both cases $V_{0}=5$. See text for details.}
    \label{fig:4}
\end{figure*}

For a detailed analysis, changes in $S_{r}$, $S_{p}$ for He (a) and O$^{6+}$ (b) are depicted in Fig.~\ref{fig:4}, where only XC-WIG results are used. On the whole, all members of He iso-electronic 
series ($Z=2-18$) show a similar trend. It may be noted that, the electron densities in $r$ and $p$ spaces are unit-normalized; moreover both the information measures ($S$ and $I$) include radial and angular 
contributions. For He in panel (a), in the limit $R\rightarrow 0$, probability density in $r$-space is quite localized and thus $S_{r}$ is small. Moving to the intermediate 
region ($R\in [0.1,10]$), with a weakening of GP, the system is increasingly governed by the term $(-\frac{Z}{r})$. Because of this strong rearrangement, the spatial distribution becomes most compact 
in this region; so $S_{r}$ reaches a minimum. When $V_{0}$ goes from 5 to 10, density profile is squeezed and thus, $S_{r}$ decreases; and hence more localization occurs. Next, in large-$R$ limit, 
the electron cloud remains practically unaltered with $R$, and $S_{r}$ appears flat again. On the contrary, $S_{p}$ spreads out, and shows an opposite nature of $S_{r}$. One also notices that, as 
$Z$ increases in O$^{6+}$ in panel (b), the qualitative trend of curves do not change; because for higher $Z$, $(-\frac{Z}{r})$ dominates and GP is effective in low regime, but very small. 
To substantiate this, clear inset plots are included to show the minima in $S_{r}$ and maxima of $S_{p}$.
%%%By increasing the impurity $Z$, the changes in entropies can be compared with He atom.
When the GP is relatively tight, the transition from dot-controlled to nucleus-controlled behavior occurs earlier and the extrema in $S_{r}, S_{p}$ shift towards lower $R$. Furthermore, their 
magnitudes are also enhanced; $S_{r}$ attains a deeper minima and $S_{p}$ reaches a higher maxima. Thus, for higher $Z$, the system reflects a stronger spatial localization and corresponding 
momentum delocalization. 
%%%\st{Such a gradual compression and expansion of electron cloud with changes in $Z$ and confinement parameters directly influence $S_{r}$.}
The contraction of charge distribution with increasing $Z$ manifests in the corresponding trends of $S_{r}$, portrayed (for $R=0.001,1, 100$) in panel (c). The effective volume of charge 
distribution reduces, because the electron cloud contracts toward nucleus. For smaller $Z$ ($Z=2-7$), the density is spread out; the GP can compress it much more strongly. So the relative change 
in density shape ($R\approx 1$) compared to the free system ($R\approx100$) is large. But as one goes to higher $Z$ ($Z>8$), $(-\frac{Z}{r})$ dominates, and confinement contributes minimally. 
Thus, $S_{r}$ becomes nearly independent of the Gaussian strength.
Overall $S_{r}$ shows decreasing trend as $Z$ increases, reflecting stronger localization. When the electron charge density contracts in $r$ space, the $p$-density distribution must broaden to 
satisfy the entropic uncertainty relation. The effect of GP induces suppression of electron-electron separation and therefore, enhances the momentum of the distribution. Therefore, it leads to an 
increase in $S_{p}$, demonstrated in panel (d). The difference between free and confined cases becomes very small at larger $Z$; so the resulting $S_{p}$ curves gradually tend to merge.

To find out the qualitative trend, we have fitted $S_r, S_p$ as functions of $Z$, leading to the following equations: 
\begin{equation}\label{srvsz}
    S_r = s_{r_0} + s_{r_1} e^{-\frac{Z}{Z_1}} 
\end{equation}
and
\begin{equation}\label{spvsz}
    S_p = s_{p_0} + s_{p_1}\left(1- e^{-\frac{Z}{Z_2}}\right).
\end{equation}
The fitting parameters corresponding to Eqs.~\eqref{srvsz} and \eqref{spvsz} are listed in Table \ref{tab:3}.

\begin{figure*}[t]
    \centering
    \begin{tabular}{cccc}
       (a)  &  (b) & (c) & (d)\\ 
\includegraphics[width=0.23\textwidth]{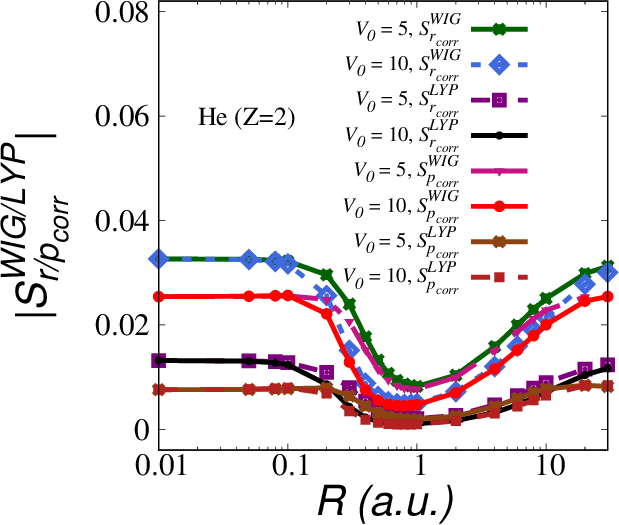}& \hspace{-0.1cm}
\includegraphics[width=0.21\textwidth]{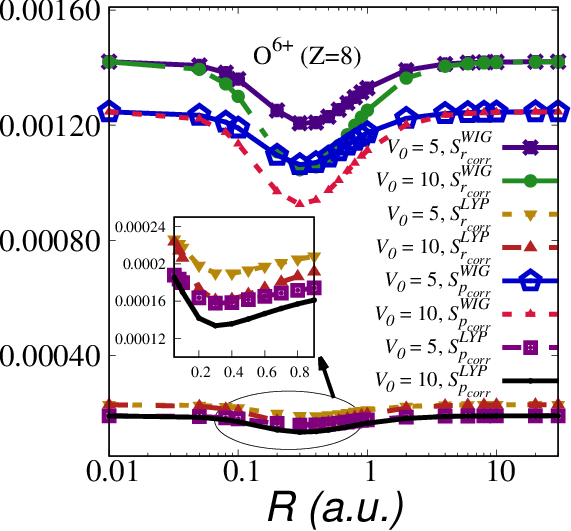}& 
\includegraphics[width=0.23\textwidth]{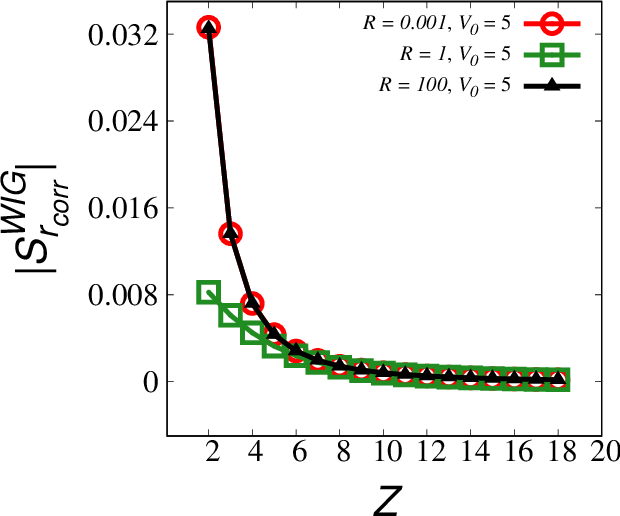}&  \hspace{-0.1cm}
\includegraphics[width=0.23\textwidth]{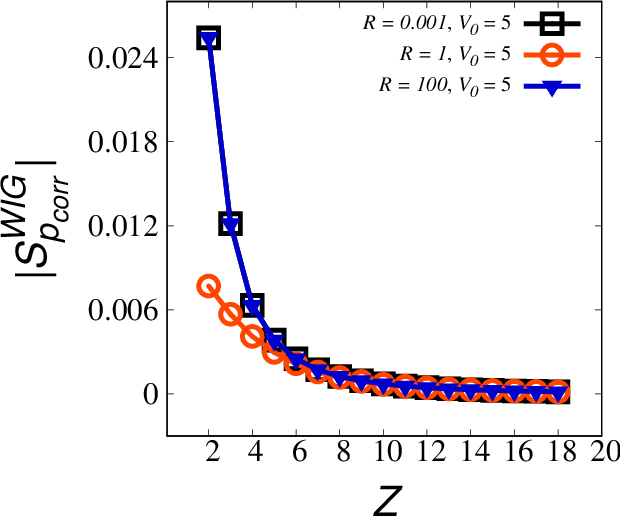}
    \end{tabular}
	\caption{Variation of $|S^{WIG/LYP}_{{\texttt{r/p}}_{\texttt{corr}}}|$ w.r.t. $R$, for $V_{0}=5$ and $10$, is shown in panels (a) and (b) for He and O$^{6+}$. Panels (c) and (d) show 
	$|S^{WIG}_{\texttt{r}_{\texttt{corr}}}|$ w.r.t. $Z$ and $|S^{WIG}_{\texttt{p}_{\texttt{corr}}}|$ w.r.t. $Z$, in He-iso-electronic series. In both cases, $V_{0}=5$. See text for details.}
    \label{fig:5}
\end{figure*}

The effect of electron correlation in terms of $S_{\texttt{r/p}}$ defines how much uncertainty or spread or delocalization arises purely from electron correlation, 
which can be qualitatively estimated through correlated $S_{\texttt{corr}}$, defined below as: 
\begin{equation}
\label{eq:27}
S_{\texttt{corr}} \approx S_{\texttt{XC}} - S_{\texttt{XX}}, 
\end{equation}
where $S_{\texttt{XC}}$ and $S_{\texttt{XX}}$ refer to $S_{\texttt{r/p}}$ estimated using XC-WIG (or XC-LYP) and XX respectively.
It may be mentioned at this point that, E$_{\texttt{XX}}$ from this work-function potential provides near-HF quality results, 
which has been confirmed in a number of previous works \cite{roy97,roy97a,roy2002}.  
A detailed variation of $S^{WIG/LYP}_{\texttt{corr}}$ has been demonstrated in Fig.~\ref{fig:5} for He and O$^{6+}$. %From the variation in panel (a), 
%%%, the variation of $S^{WIG/LYP}_{\texttt{corr}}$ as a function of $R$ for the ground state of He atom is shown. Evidently, 
It is observed that for He atom in panel (a), in the limit $R \rightarrow 0$, the system is equivalent to dot size under the influence of GP. Consequently, both electrons are strongly 
constrained, which shows that electron delocalization is almost saturated. Thus, in this limit, the nature of the curve is almost flat. 
But, in the intermediate regime $R\in[0.1,10]$, electrons have combined effect of GP and ($-\frac{Z}{r}$) potential, where electronic charge distribution is more ordered, which reflects localization 
and thereby having a minimum. On the other hand, as $R\rightarrow \infty$, the effect of GP gradually diminishes and the system tends to behave as a free system. In this region, 
electron correlation again gets saturated, and $S_{\texttt{corr}}$ converges to a smooth or nearly constant behavior. 

Note that, as $Z$ increases, the magnitude of $S^{WIG/LYP}_{{\texttt{r/p}}_{\texttt{corr}}}$ becomes smaller, but its overall trend remains same as clearly depicted in panel (b). 
The pattern behavior of $S^{WIG}_{\texttt{r}_{\text{corr}}}$ and $S^{WIG}_{\texttt{p}_{\text{corr}}}$ w.r.t. to $Z$ in He-like ions have been revealed in panels (c) and (d). These two plots exhibit 
quite similar trends, 
%%%From an analysis on the dependence of $S^{WIG}_{\texttt{r}_{\text{corr}}}$ and $S^{WIG}_{\texttt{p}_{\text{corr}}}$ on $Z$ have been shown 
%%%in Fig.~\ref{fit:ScorrvsZ}. The fitted data (solid lines) and our evaluated values (symbols) are depicted in the same figure. We note that $S_{\text{r,corr}}$ 
satisfying the following fitted equation:
\begin{equation}\label{srcorrvsz}
    S_{\text{r,corr}} = s_{0}^{\text{r,corr}} + s_{1}^{\text{r,corr}} \left(1-e^{-\frac{Z}{Z_3}}\right), 
\end{equation}
while $S_{\text{p,corr}}$ obeys
\begin{equation}\label{spcorrvsz}
S_{\text{p,corr}} = s_{0}^{\text{p,corr}} + s_{1}^{\text{p,corr}} \left(1-e^{-\frac{Z}{Z_4}}\right). 
\end{equation}
The respective fitting parameters in above equations are given in Table~\ref{tab:3} along with the reduced $\chi^2$ values.

\subsection{Fisher information entropy} \label{sec:fisher}
Next, we turn our focus to $I$, that measures sharpness or concentration of an electron charge density, as it includes the gradient of density. In other words, it provides a local measure of 
charge density; thus it measures information point-to-point of the electron cloud distribution. In $r$-space, it is written as,  
%%%in terms of gradient of $\rho(\vec{r})$ 
\begin{equation}
\label{eq:28}
I_r = \int \frac{\left| \nabla \rho(\vec{r}) \right|^{2}}{\rho(\vec{r})} d{\vec{r}}.
\end{equation}
It is a slightly difficult task to evaluate $I_{r}$ at the origin. To address this, we have added the extrapolation to the density at the origin. The polynomial extrapolation is given explicitly by 
Lagrange's classical formula \cite{septianto}, 
\begin{equation}
\label{eq:29}
P(x)=\sum_{i=1}^k Y_i l_i(x); ~~
%\end{equation} 
%with
%\begin{eqnarray}
%\label{eq:30}
    l_i(x) = \prod_{\substack{j=1 \\ j\neq i}}^{k}
\frac{x - x_j}{x_i - x_j}.
\end{equation}
The known points defined here are: ($x_{i}$=$r_{i}$, $Y_{i}$ = $C_{i}$) with $r_{i}$ = $f(x_{i})$, $C_{i}$=$\psi_{l}(r_{i})r_{i}^{-l-1}$.

\begin{figure*}[t]
    \centering
    \begin{tabular}{ccc}
       (a)  &  (b)  & (c)\\
\includegraphics[width=0.32\textwidth]{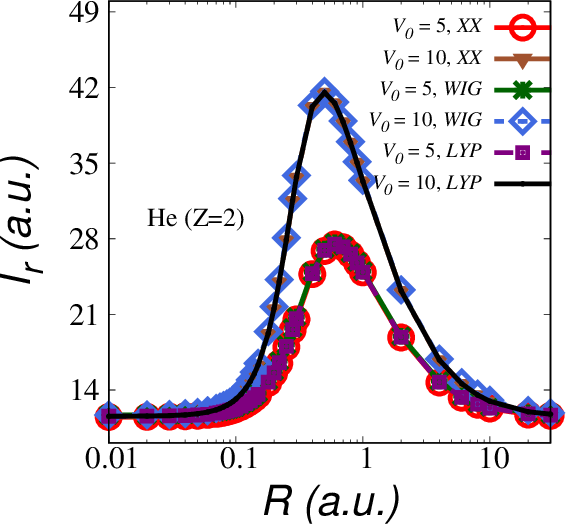}& 
\includegraphics[width=0.29\textwidth]{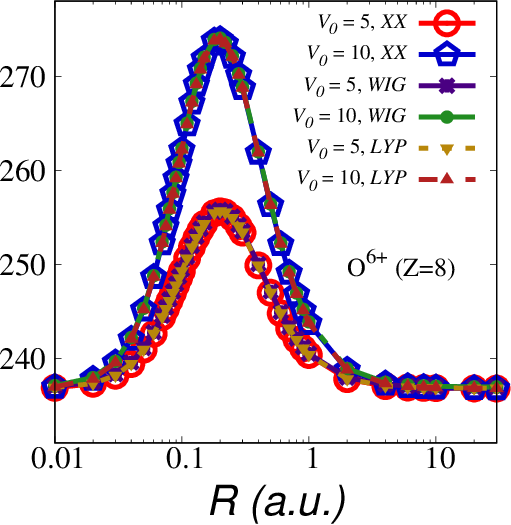}&
\includegraphics[width=0.305\textwidth]{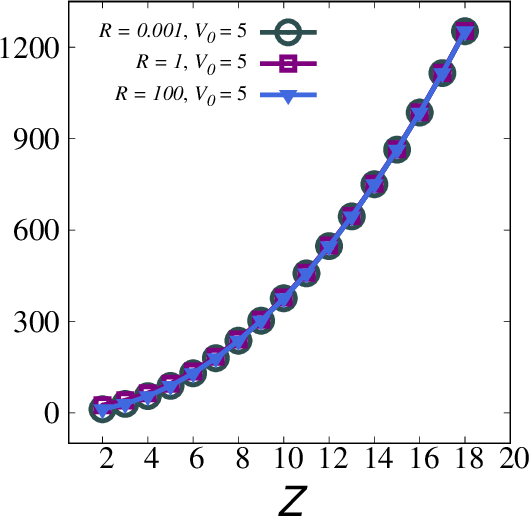}
\end{tabular}
	\caption{Variation of $I_{r}$ w.r.t. $R$ with XX, XC-WIG and XC-LYP functionals, for $V_{0}=5$ and $10$ in panels (a) and (b) He and O$^{6+}$. Panel (c) shows $I_{r}$ w.r.t. $Z$, in He-like 
	ions, with XC-WIG, $V_{0}=5$. See text for details.}
    \label{fig:6}
\end{figure*}

Figure~\ref{fig:6} depicts $I_{r}$ w.r.t. $R$ and $V_{0}$ plots for He in GP. Panels (a) and (b) correspond to He and O$^{6+}$ ion; for XX, XC-WIG and XC-LYP functionals having two 
$V_0$ (5, 10). For a given $R$ and $V_0$, the functionals tend to produce similar entropies. When $R \lesssim 0.1$, GP is too narrow such that it does not appreciably modify the density. 
Thus, in this region, He behaves as the free system, which is reflected from $I_{r}$ value. At the other extreme, when $R\gtrsim 20$, GP becomes extremely broad, and hence the effect of 
confinement almost vanishes. Electrons in this case, too experience almost no additional pull; so again $I_{r}$ returns to free-atom limit. The interesting behavior can be found in the 
intermediate region, where GP is neither too narrow nor too broad. A clear maximum appears around at $R \approx 0.6$, when $V_{0}=5$, at which the effect of GP is strong enough to pull 
the electrons inward, while they still try to spread out because of the mutual repulsion between each other. Now, for the region $ 0.1 \lesssim R\lesssim 0.6$, 
the density gradient increases sharply. But for $R\gtrsim0.6$, the potential is not tight enough to create sharply rising slopes in density. Therefore 
$I_{r}$ begins to diminish monotonically.

%%%%%%%%%%%%%%%%%%%%%%%%%%%%%%%%%%%%%%%%%%%%%%%%%%%%%%%%%%%%%%%%%%%%%%%%%%%%%%%%%%%%%%%%%%
For ions in He-like ($Z=2-18$) systems, the overall pattern remains quite similar as that for He. However, the $I_{r}$ profile shifts toward the origin or low value of $R$. 
This occurs because the charge distribution in higher $Z$ is already strongly localized around the nucleus, but inclusion of GP leads to a confinement effect that enhances the 
localization and strengthens radial gradients, which occurs at a lower $R$, as revealed from panel (b). This leads to the maxima of $I_{r}$ in a lower $R$. For instance, $I_{r}$ 
reaches its maximum at $R\approx 0.2$ for O$^{6+}$, when $V_0=5$. But when $V_{0}$ is increased from 5 to 10, confinement becomes much stronger, and 
this directly enhances localization of charge distribution. A large $V_0$ produces a larger gradient, and since $I_{r}$ is a gradient-based measure, it produces a higher peak in $I_{r}$. 
%while the same occurs at $R\approx0.13$ and $R\approx0.09$ for Al$^{13+}$ and Ar$^{16+}$. 

Now, to examine the explicit dependence of $I_{r}$ on $Z$, panel (c) displays the calculated behavior for $V_0=5$ with XC-WIG at three representative $R$ (0.001, 1, 100). As seen, 
it provides a fundamentally different perspective on electron distribution, compared to $S_{r}$. A simple fitting for $I_r$ vs. $Z$ leads to the following parabolic equation:
\begin{equation}\label{irvsz}
    I_r=I_{r_0}+ I_{r_1} Z^2,
\end{equation}
where $I_{r_0}$, $I_{r_1}$ are fitting parameters, listed in Table~\ref{tab:3}. 

\begin{figure*}[!th]
    \centering
    \begin{tabular}{ccc}
       (a)  &  (b)  & (c)\\ \hspace{-1cm}
\includegraphics[width=0.32\textwidth]{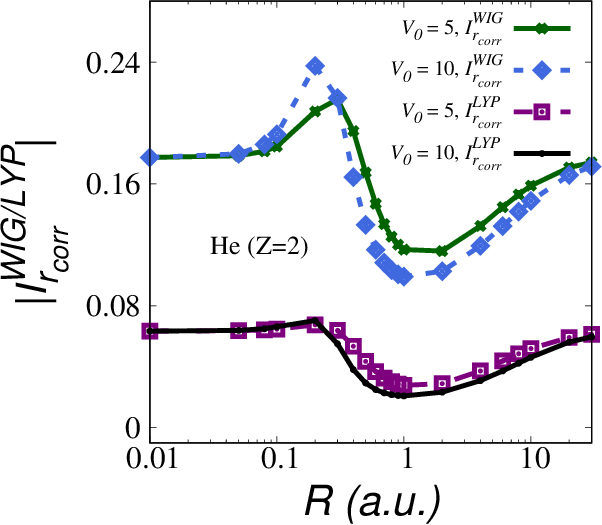}& 
\includegraphics[width=0.27\textwidth]{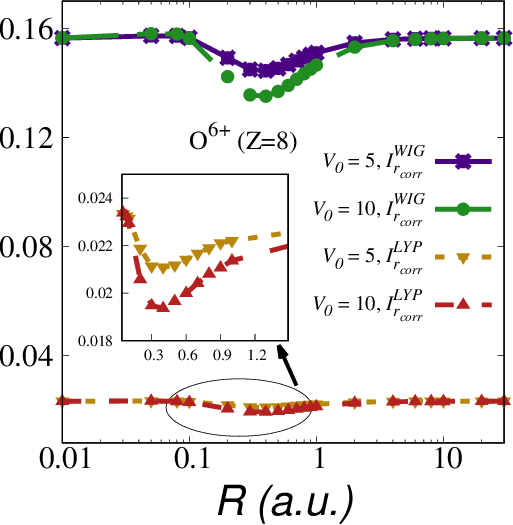}& 
%(c) & (d)\\
\includegraphics[width=0.33\textwidth]{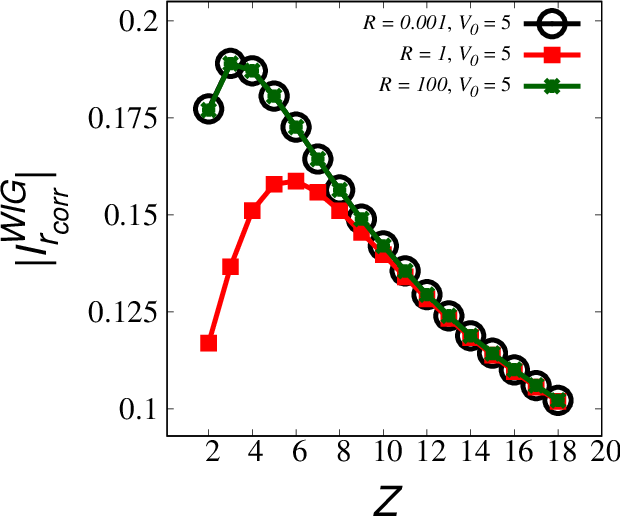}
    \end{tabular}
    \caption{Variation of $I_{\text{r}_\text{corr}}$ for XC-WIG, and XC-LYP w.r.t. $R$, for $V_{0}=5$ and $10$ in panel (a) He ($Z = 2$), in (b) O$^{6+}$ ($Z = 8$), and in (c) variation of $I^{WIG}_{\text{r}_\text{corr}}$ w.r.t. $Z$ with XC-WIG and $V_{0}=5$. See text for details.}
    \label{fig:7}
\end{figure*}

Next, we study correlated Fisher information $I_\texttt{corr}$, defined analogous to that as for $S_\texttt{corr}$ and $E_{\texttt{corr}}$ as, 
\begin{equation}
\label{eq:31}
I_{\texttt{corr}} \approx I_{\texttt{XC}} - I_{\texttt{XX}}, 
\end{equation}
where $I_{\texttt{XC}}$ and $ I_{\texttt{XX}}$ refer to $I_{r}$ with XC-WIG (XC-LYP) and XX respectively. These variations are recorded in Fig.~\ref{fig:7}--panels (a) and (b) for 
He and O$^{6+}$. One finds that the overall nature of $I^{WIG/LYP}_{r_{\texttt{corr}}}$ for all systems is quite reminiscent to that of $S^{WIG/LYP}_{r_\texttt{corr}}$, except in 
the neighborhood of $R\rightarrow 0$, where it behaves differently. The obtained particular behavior of He atom is expected, since the electronic charge distribution is fairly 
delocalized in space. Thus, $I_{r}$ shows a non-monotonic behavior in the range of $0.05$ to $20$ for $R$. When $Z$ increases, e.g., O$^{6+}$ in panel (b), width of such a region also reduces and 
shifts toward a lower $R$, due to the dense profile of density near nucleus. To illustrate this, the inset plot shows a magnified view of $I^{LYP}_{r_{\texttt{corr}}}$. 

The trend for $I^{WIG}_{\text{r}_\text{corr}}$ as a function of $Z$ is displayed in panel (c) for He-like ions. Three $R$ (0.001, 1, 100) values are considered for a fixed 
$V_0=5$ for XC-WIG. The nature of $I^{WIG}_{\text{r}_\text{corr}}$ is described by the following fitted equation, 
\begin{equation}\label{eq:35}
    I^{WIG}_{\text{r}_\text{corr}}  = \sum_{i=0}^3 I^{\text{r}_\text{corr}}_{i}.
\end{equation}
The fitting parameters are reported in Table~\ref{tab:3}, along with the $\chi^2$ reduced values.

\subsection{Fisher-Shannon information plane} \label{sec:fisher-shannon}
\begin{figure}[b]
    \centering
    \begin{tabular}{c}      
\includegraphics[width=0.35\textwidth]{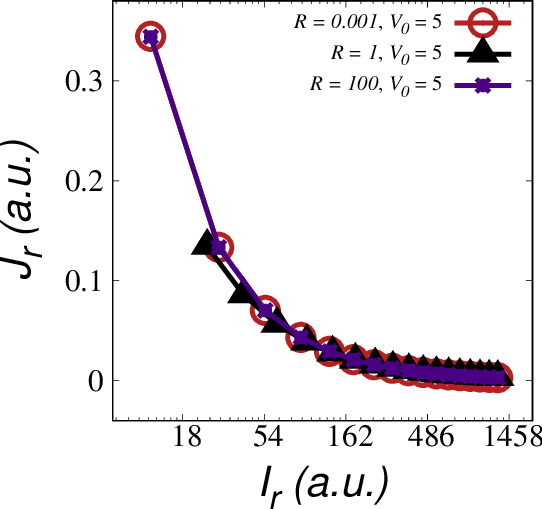}
    \end{tabular}
    \caption{The Fisher-Shannon information plane for He-like $Z=2-18$ systems with $V_{0}$=5. See text for details.}
    \label{fig:8}
\end{figure}

When a system is kept in an external environment, its density is significantly modified. The Fisher-Shannon plane is used conveniently to simultaneously examine the disorder 
(global) $S$ and order (local) $I$ of a system, thereby enabling the study of correlation effects \cite{dehesa,vignat,angulo}. In $r$-space, the plane is defined as, 
\begin{equation}
    P_{r}=J_{r}I_{r},
\end{equation}
where $J_{r}$ defines the Shannon entropy power \cite{dembo91}, 
\begin{equation}
J_{r}=1/(2\pi e)e^{2S_{r}/3}.    
\end{equation}
The nature of $J_{r}$ and $I_{r}$ can be observed in Fig.~\ref{fig:8}, where it is plotted as a function of $Z$; we have considered only XC-WIG with $V_{0}=5$. 
The pattern for three values of $R$ has been presented. Every point from left to right signifies the values of $I_r$ and $J_r$ for $Z$ varying from 2 to 18. When viewed on this 
plane, the local and global measure pair can be observed simultaneously, which captures the overall delocalization (spread) and localization (how sharply density changes in space). The X-axis has been taken in logarithmic scale. The first point at far top left end of the curve (denoted by $\bigcirc$ and $\star$) corresponds to $R=0.001$ and $R=100$, for He. 
\begin{figure*}[t]
    \centering
    \begin{tabular}{ccc}
       (a)  &  (b) & (c)\\ \hspace{-0.5cm}
\includegraphics[width=0.30\textwidth]{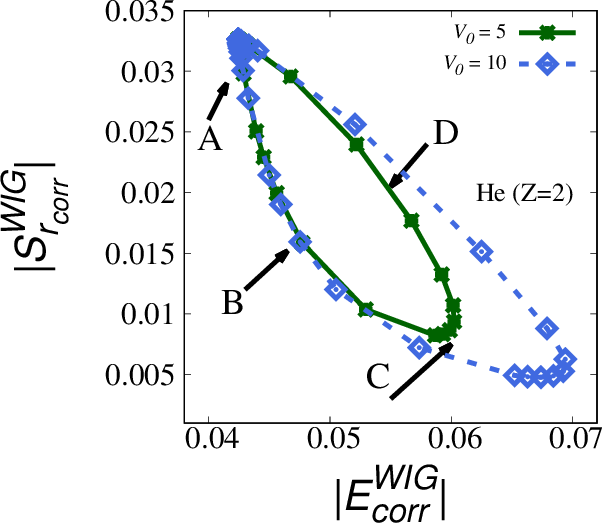}&
\includegraphics[width=0.30\textwidth]{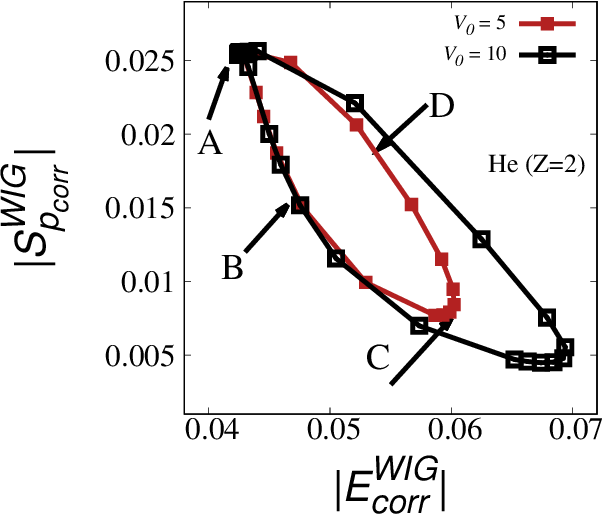}&
\includegraphics[width=0.30\textwidth]{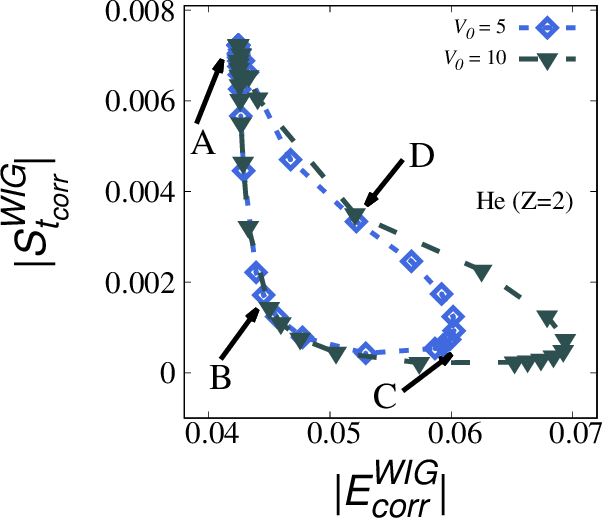} 
\end{tabular}
\caption{Variation of $S^{WIG}_{\texttt{r}_{\texttt{corr}}}$ w.r.t. $E^{WIG}_{\texttt{corr}}$ in (a), $S^{WIG}_{\texttt{p}_{\texttt{corr}}}$ w.r.t. $E^{WIG}_{\texttt{corr}}$ in (b) 
and $S^{WIG}_{\texttt{t}_{\texttt{corr}}}$ w.r.t. $E^{WIG}_{\texttt{corr}}$ in (c). These correspond to $V_{0}=5$ and $10$ with XC-WIG functional. See text for details.}
    \label{fig:9}
\end{figure*}
Here $I_{r}$ is small; the density is relatively smooth and extended, so it provides a large value of $J_{r}$. But when we tune $R$ in the intermediate region (like $R=1$), considerable change in density takes place, as evidenced from the respective $J_r$ and $I_r$ values, denoted by the first $\blacktriangle$ at the left side. Thus it is discerned that at $R=0.001$ (low) and $R=100$ (high), the system behaves like a free atom, while confinement effect becomes more pronounced in the intermediate range. The $J_{r}$ remains comparatively below the free limit, while $I_{r}$ increases depicting the localization because now due to the GP, electron density is sharply peaked near the nucleus. Similarly, the second point of the curve (represented by $\bigcirc$ and $\star$ symbol) signifies low and high regions of $R$ for Li$^+$, which almost coincides with first point for He, in intermediate ($R=1$) region. Once again the intermediate $R$ region for Li$^+$ ($\blacktriangle$) is also little bit shifted right. This phenomenon continues for other higher members of the series in a similar stair-case fashion. Eventually, for higher $Z$, since the density is already tightly bound near the nucleus (due to dominant effect of attraction term), after a certain $Z$ ($\approx7$), the influence of GP starts to diminish. Thus, the three $R$ curves tend to merge into a single trajectory. 

\subsection{Collin's conjecture}\label{sec:collins}
This subsection is based on the original works of Collins \cite{collins} and researchers thereafter \cite{smith,site,gallegos,grassi,flores,march,rodolfo}. 
The primary objective is to investigate the interrelation between $S_{\texttt{corr}}$ and $E_{\texttt{corr}}$ under the GP. 
The joint analysis provides a multidimensional characterization of electron correlation that goes beyond traditional energy-based descriptions. 
A detailed study of $S_{\texttt{corr}}$ has been reported for endohedrally confined atomic systems\cite{saha1} and for weakly homogeneous electron gas via investigating helium iso-electronic ions \cite{guevara}. 
In Secs.~\ref{sec:energy} and \ref{sec:Shannon}, the behavioral pattern of $E_{\texttt{corr}}$ and $S_{\texttt{corr}}$ with $R$ 
has been dealt with individually. Now it would be interesting to consider changes of $S_{\texttt{corr}}$ and $E_{\texttt{corr}}$ as functions of $R$  
and $V_0$ simultaneously. Similar to $E_{\texttt{corr}}$ and $S_{\texttt{corr}}$, we define following three equivalent quantities, such as: (a) $S_{\texttt{r}_{\texttt{corr}}}$ in $r$ space (b) 
$S_{\texttt{p}_{\texttt{corr}}}$ in $p$ space (c) $S_{\texttt{t}_{\texttt{corr}}}$ as the total correlated quantity in $r$ and $p$ spaces jointly. These are calculated 
for both functionals; but for sake of simplicity, only results for XC-WIG are shown in Fig.~\ref{fig:9}. For He, panels (a), (b), (c) of the figure 
record some interesting features. It is worth mentioning that every point signifies the correlation contribution for a specific $R$ and $V_0$. In each plot, 
the inner and outer loops indicate results for $V_{0}=5$ and 10 respectively. In panel (a), in the small regime of $R$ or strong GP confinement (AB segment), as 
one moves from point A to B, $S^{WIG}_{\texttt{r}_{\texttt{corr}}}$ drops down, while $E^{WIG}_\texttt{corr}$ slowly increases. But in this strong confinement
region, overall $S^{WIG}_{\texttt{r}_{\texttt{corr}}}$ is high and $E^{WIG}_\texttt{corr}$ low, due to a lesser effective space between two electrons. The 
most significant physics emerges in intermediate region of $R$ (BCD portion of curve), where the underlying GP confinement and ($-\frac{Z}{r}$) 
compete with each other. The electrons are less confined and have appropriate space to move. Thus, a change in density leads to stronger correlation and 
system localization. In BC path, $E^{WIG}_{\texttt{corr}}$ increases while $S^{WIG}_{\texttt{r}_{\texttt{corr}}}$ lowers. With increase in $R$ in 
path CD, one sees that the correlated energy declines with small change, but $S^{WIG}_{\texttt{r}_{\texttt{corr}}}$ sharply increases. As $R$ increases 
further in CD section, the correlated energy falls slightly, but correlated entropy enhances reasonably. Even further  enhancement in $R$ in path DA leads to 
both the correlated quantities returning back to the point when $R$ was very small, thus completing an overall loop structure. For $V_0=10$, the system becomes 
more localized providing a larger loop. A similar loop structure is observed for panels (b) and (c) depicting $S^{WIG}_{\texttt{p}_{\texttt{corr}}}$ and 
$S^{WIG}_{\texttt{t}_{\texttt{corr}}}$ . Also, the magnitude of $S_{\texttt{corr}}$ decreases 
when one goes from panel (a) to (b) to (c). For higher $Z$ ions in the series, impurity retains a similar trend but position in 
$S^{WIG}_{\texttt{r}_{\texttt{corr}}}-E^{WIG}_{\texttt{corr}}$ plane shifts systematically. The magnitude of $E^{WIG}_{\texttt{corr}}$ increases while $S^{WIG}_{\texttt{r}_{\texttt{corr}}}$ decreases.

\section{Conclusion}\label{sec:conc}
Energetic and information-theoretic analysis have been made for atoms/ions confined in a GP within KS DFT framework. This is achieved 
by means of an accurate work-function exchange along with two approximate correlation energy functionals. Numerical solution is made with 
the help of GPS method. By tuning the depth and width of potential, some interesting features in energy for H, He-iso-electronic series 
as well as Li, Be are observed. Apart from ground state, some low-lying states are also considered. Some notable features are:
(i) the most significant energy changes take place in the intermediate range of $R$ (0.01-10)
(ii) for the He-isoelectronic series, $S_{r}$ show minima whereas $S_{p}$ and $I_{r}$ reveal maxima 
(iii) the contribution of correlation in higher $Z$ is less dramatic than that in lower $Z$
(iv) effect of GP on structural properties is minimal beyond $Z \approx 7$
(v) an interesting loop structure has been observed for all systems in $S^{WIG}_{\texttt{r}_{\texttt{corr}}}$, $S^{WIG}_{\texttt{p}_{\texttt{corr}}}$, and 
$S^{WIG}_{\texttt{t}_{\texttt{corr}}}$ w.r.t. $E^{WIG}_{\texttt{corr}}$ with respect to changes in $R$ and $V_{0}$. It exhibits a distinct behavior in presence 
of GP confinement compared to the respective free system.
(vi) for all the following quantities, $E_{\texttt{corr}}$, $S_{\texttt{r}_{\texttt{corr}}}$, $S_{\texttt{p}_{\texttt{corr}}}$ and $I_{\texttt{r}_{\texttt{corr}}}$, 
it is found that XC-WIG > XC-LYP
(vii) some simple fitting equations are given for energy as well as information related quantities.
This study under GP will likely motivate future work using other model confining potentials, which may reveal novel features.  
The study may be extended to complexity measures, entanglement measures as well. 

\subsection*{Acknowledgement}
RA thanks CSIR-UGC, Govt. of India, for the financial support. The postdoctoral stay of the SM is supported by the DGAPA-UNAM Postdoctoral Scholarship Program. AKR acknowledges partial support from SERB, New Delhi (sanction order CRG/2023/004463).

\bibliographystyle{apsrev4-1}
\bibliography{biblio}
\end{document}